\documentclass[aps,prd,amsmath,amssymb,preprintnumbers,superscriptaddress,a4paper,11pt,nofootinbib]{revtex4-1}
\pdfoutput=1
\usepackage{amsthm}
\usepackage{graphicx}
\usepackage[ margin=5pt, font=small,labelfont=bf, justification=raggedright]{caption}
\usepackage{url}
\usepackage[bookmarks, pagebackref=false]{hyperref}
\usepackage{color}
	\definecolor{rossoCP3}{cmyk}{0,.88,.77,.40}
		\definecolor{graa}{rgb}{0.8,0.8,0.8}
		\definecolor{blaa}{rgb}{0.2,0.2,0.6}
		\hypersetup{
			colorlinks,
			bookmarksopen,
			bookmarksnumbered,
			citecolor=blaa, 		
			linkcolor=rossoCP3,	
			urlcolor=rossoCP3,			
			}
\usepackage{dcolumn}
\usepackage{bm}
\usepackage{bbm}
\usepackage{pxfonts}

\usepackage{amsmath}
\usepackage{amsfonts}

\usepackage{dcolumn}
\usepackage{bm}
\usepackage{amssymb}
\usepackage{latexsym}
\usepackage{color}

\usepackage{epsfig}
\usepackage{placeins}

\usepackage{youngtab}
\usepackage{slashed}
\usepackage{subfig}

\newcommand{\be}{\begin{equation}}
\newcommand{\ee}{\end{equation}}
\newcommand{\bea}{\begin{eqnarray}}
\newcommand{\eea}{\end{eqnarray}}

\DeclareRobustCommand{\SkipTocEntry}[5]{}

\begin{document}

\title{\Large  Dynamical Origin of the Electroweak Scale and the 125 GeV Scalar}

\author{Stefano Di Chiara}
\author{Roshan Foadi}
\affiliation{Department of Physics, University of Jyv\"askyl\"a, P.O. Box 35, 
FI-40014, University of Jyv\"askyl\"a, Finland}
\affiliation{Helsinki Institute of Physics, P.O. Box 64,
FI-000140, University of Helsinki, Finland}
\author{Kimmo Tuominen}
\affiliation{Department of Physics, University of Helsinki, P.O. Box 64,
FI-000140, University of Helsinki, Finland}
\affiliation{Helsinki Institute of Physics, P.O. Box 64,
FI-000140, University of Helsinki, Finland}
\author{Sara T\"ahtinen}
\affiliation{Department of Physics, University of Jyv\"askyl\"a, P.O. Box 35, 
FI-40014, University of Jyv\"askyl\"a, Finland}
\affiliation{Helsinki Institute of Physics, P.O. Box 64,
FI-000140, University of Helsinki, Finland}

\preprint{HIP-2014-39/TH}

\begin{abstract}
We consider a fully dynamical origin for the masses of weak gauge bosons and heavy quarks of the Standard Model. Electroweak symmetry breaking and the gauge boson masses arise from new strong dynamics, which leads to the appearance of a composite scalar in the spectrum of excitations. In order to generate mass for the Standard Model fermions, we consider extended gauge dynamics, effectively represented by four fermion interactions at presently accessible energies. By systematically treating these interactions, we show that they lead to a large reduction of the mass of the scalar resonance. Therefore, interpreting the scalar as the recently observed 125 GeV state, implies that the mass originating solely from new strong dynamics can be much heavier, {\em {\em i.e.}}  of the order of 1 TeV. In addition to reducing the mass of the scalar resonance, we show that the four-fermion interactions allow for contributions to the oblique corrections in agreement with the experimental constraints. The couplings of the scalar resonance with the Standard Model gauge bosons and fermions are evaluated, and found to be compatible with the current LHC results. Additional new resonances are expected to be heavy, with masses of the order of a few TeVs, and hence accessible in future experiments.
\end{abstract}

\maketitle

\tableofcontents

\section{Introduction}
\label{Sec:intro}

The discovery of the Higgs boson at the Large Hadron Collider (LHC) \cite{Aad:2012tfa, Chatrchyan:2012ufa} established 
the Standard Model (SM) as an accurate description of elementary particle interactions \cite{Giardino:2013bma,Azatov:2012bz,Alanne:2013dra}.  However, it is know that the SM is incomplete: For example, the SM itself does not provide any clue towards understanding the generational structure and mass patterns of the matter fields. Furthermore, understanding the origin of dark matter or the baryon-antibaryon asymmetry continue to provide motivation for searches of viable beyond-the-Standard-Model (BSM) scenarios. 

So far the LHC has shown no sign of new particles typically predicted by various BSM setups, such as Technicolor (TC)  and its variants (see \cite{Hill:2002ap, Sannino:2009za} for review). Furthermore, the lightest resonance in a model of dynamical electroweak symmetry breaking is naturally expected to be much heavier than $M_H\simeq 125$ GeV \cite{Foadi:2007ue}. These premature concerns rest on treating new strong dynamics in isolation, {\em i.e.} without taking the interaction with the SM fields into account. It is known that a light scalar can arise from approximate global symmetries, as in models where the Higgs is a pseudo Goldstone boson associated with chiral symmetry \cite{Kaplan:1983fs, Kaplan:1983sm, Cacciapaglia:2014uja} or scale invariance \cite{Yamawaki:1985zg, Bando:1986bg, Sannino:2004qp,Dietrich:2005jn,Chacko:2012sy}.  Only recently it has been realised that also with QCD-like TC dynamics the scalar particle can become light because of loop corrections originating from extended sectors, which are always required in TC models to account for the generation of fermion masses. In \cite{Foadi:2007ue} a preliminary analysis, using simply SM-like Yukawa couplings to parametrize the effects from the coupling with the top quark, was carried out to point out this effect. In \cite{DiChiara:2014gsa} this effect was investigated in a fully dynamical model setup of simple extended technicolor (ETC). Within this model, a computation in the large-$N$ limit was carried out, where $N$ is the dimension of the technifermion representation under the TC gauge group. It was then possible to rigorously demonstrate a large reduction of the scalar mass from the value arising solely from new strong dynamics. The amount of fine tuning involved is on the tolerable level of a few per cent \cite{DiChiara:2014gsa}. However, the model considered in \cite{DiChiara:2014gsa} was simple and devised only to illustrate this effect, and it could not be used for a realistic description of the origin of all mass scales of the SM. 

In this paper we present a necessary further development of the model framework described above. We use a chiral fermion model, similar to the Nambu--Jona-Lasinio model (NLJ), to account for TC dynamics, and augment it with a whole set of four-fermion operators, low-energy remnants of ETC interactions. We show that the mechanism featured in \cite{DiChiara:2014gsa} for the reduction of the scalar mass also works in this case, and that the effective couplings of the composite Higgs particle with the SM particles are very close to the SM-Higgs couplings, and hence compatible with the LHC data\footnote{See also \cite{Espriu:2013fia} for a related study.}.  We also compute the oblique corrections and demonstrate the viability of the model with respect to the electroweak precision data. One of our robust and generic findings within this framework is that in order to reduce the Higgs mass from values near 1 TeV, natural for new strong dynamics, to 125 GeV, the ETC interactions must be strongly coupled. However, we only consider scenarios in which the ETC interactions, although strong, are not strong enough to generate fermion condensation. Therefore, we complement the analysis of \cite{Lane:2014vca}, where a model with strong ETC dynamics and weak TC interactions was considered.

Model building of the full gauge dynamics required by ETC theories is challenging \cite{Appelquist:2003hn}. Our effective theory, formulated in terms of four fermion couplings, and taking into account only the third generation quarks, can hopefully be seen as a stepping stone towards more complete dynamical theories of flavour. There exists lots of earlier work using NJL-like models to describe dynamical electroweak symmetry breaking and the associated Higgs physics, see {\em e.g.}. \cite{Miransky:1988xi,Miransky:1989ds,Bardeen:1989ds, Dobrescu:1997nm,Chivukula:1998wd, He:2001fz, Fukano:2012qx,Fukano:2013kia,Geller:2013dla}. There is also a large lattice program motivated by applications to BSM physics and aimed at studying strong dynamics in isolation \cite{Hietanen:2008mr, Hietanen:2009az, Karavirta:2011zg, Appelquist:2011dp, Appelquist:2013pqa, DeGrand:2013uha, Fodor:2012ty, Fodor:2014pqa, Aoki:2013zsa, Aoki:2014oha}. Our analysis should be applicable in refining the phenomenological interpretation of the lattice results.

The paper is organised as follows. In sections \ref{Sec:chiralTQ} and \ref{Sec:ETC} we introduce the effective description of the strong TC dynamics, and the interactions arising from the ETC theory, in terms of a chiral-techniquark model augmented with four fermion interactions. In section. \ref{Sec:Cutoff} we show how confinement and cutoff are realised in the model when smearing the momentum integrals with a mass distribution density for the techniquarks. In section \ref{Sec:Masses} we demonstrate how the fundamental SM fields acquire mass dynamically, whereas in section \ref{Sec:scalarmass} we prove that a strongly-coupled yet subcritical ETC theory may lead to a large reduction of the mass of the lightest scalar resonance from values near 1 TeV to 125 GeV. In section \ref{Sec:couplings} we compute the coupling of the scalar resonance with the fundamental SM fields, and in section \ref{Sec:ewparams} we evaluate the oblique electroweak paramaters. In section \ref{Sec:Results} we present the numerical results of our analysis for two different TC theories, and compare with precision data as well as LHC results. Finally, in section \ref{Sec:checkout} we conclude and discuss the further prospects.
\section{Chiral-techniquark Lagrangian}
\label{Sec:chiralTQ}
We focus on TC theories featuring one colorless weak technidoublet, $Q\equiv(U,D)$, in the complex $N$-dimensional representation of the TC gauge group. We assume that there are no additional weak doublets. Therefore, in order to avoid the topological Witten anomaly, $N$ must be an even number. Since the spinorial representation is not complex, we must have $N=4,6,8,\dots$. Cancellation of the standard gauge anomalies, and requiring the electromagnetic gauge group to remain unbroken, impose the hypercharge assignments
\begin{equation}
Y_{Q_L} = 0\ ,\quad
Y_{U_R} = \frac{1}{2}\ , \quad
Y_{D_R} = -\frac{1}{2}\ .
\end{equation}
In the limit of zero electroweak gauge couplings, the techniquark kinetic terms feature a global $SU(2)_L\times SU(2)_R$ chiral symmetry, which is dynamically broken by the TC force to $SU(2)_V$. The lightest states, and the only ones that we include in our analysis, are therefore expected to be the massless technipion triplet -- which upon electroweak gauging become the longitudinal component of the $W$ and $Z$ boson -- and a scalar singlet $H$, which will be identified with the Higgs particle. In order to model TC dynamics, we employ a chiral-techniquark Lagrangian featuring both constituent techniquarks and resonances. This reads
\begin{eqnarray}
{\cal L} &=& {\cal L}_{\overline{\rm SM}} + \overline{Q}_L i\slashed{D} Q_L + \overline{U}_R i\slashed{D} U_R + \overline{D}_R i\slashed{D} D_R 
-M_Q\left(1+\frac{y}{M_Q}H+\cdots\right)\left(\overline{Q}_L \Sigma Q_R +\overline{Q}_R \Sigma^\dagger Q_L\right)  \nonumber \\
&-&\frac{M^2}{2}H^2+\cdots +{\cal L}_{\rm ETC}\ ,
\label{Eq:L}
\end{eqnarray}
where ${\cal L}_{\overline{\rm SM}} $ is the SM Lagrangian without the terms containing the Higgs doublet, the covariant derivatives are with respect to the electroweak gauge fields, the ellipses denote higher-order terms in $H$, and the TC gauge indices have been suppressed from the techniquark fields. The field $\Sigma$ is the standard non-linear sigma-model field,
\begin{equation}
\Sigma \equiv \exp \frac{2\, i\, \Pi^i\, T^i}{v}\ ,
\end{equation}
where $\Pi^i$ is the technipion triplet, $2T^i$ are the Pauli matrices, and $v$ is the vacuum expectation value. The composite nature of the $H$ and $\Pi^i$ fields in (\ref{Eq:L}) is manifest  because the corresponding kinetic terms are absent. These are generated radiatively, and vanish at some large compositeness scale. Finally, ${\cal L}_{\rm ETC}$ contain four-fermion operators which are obtained by integrating out the heavy ETC gauge bosons. These are considered in more detail in the next section. 

\section{Four-fermion operators from ETC}
\label{Sec:ETC}
In ${\cal L}_{\rm ETC}$ we only consider four-fermion operators containing the techniquark doublet and the top-bottom doublet $q\equiv (t,b)$. In fact, operators built out of lighter SM fermions are expected to arise from exchanges of very heavy ETC bosons, and are therefore highly suppressed at the electroweak scale. We focus on ETC theories in which left-handed and right-handed fields belong to different representations, and classify the four-fermion operators according to the quantum numbers of the exchanged ETC gauge bosons. We assume that there is only one ETC gauge boson with a given set of quantum numbers, and that, under the TC and QCD gauge groups, the ETC bosons are either singlets or $N$- and $N_c$- multiplets, respectively. This categorizes the ETC bosons into five distinct classes which we call A, B, C, D, and E: The classes A and B correspond to bosons which are TC and QCD singlet with hypercharge $Y=0$ (for class A) and $Y=1$ (for class B). The classes C, D and E consist of bosons which are $N$- and $N_c$-multiplets of TC and QCD, respectively, with hypercharge $Y=1/6$ for class C, $Y=5/6$ for class D and $Y=7/6$ for class E.

Below the ETC scale the ETC gauge bosons $G_\mu$ are heavy with masses ${\cal M}_G$. Integrating out the heavy bosons leads to effective four fermion interactions. Generally, the relevant terms in the fundamental Lagrangian are of the form
\be
{\cal L}_{\rm{ETC}}^G\sim g_{XX^\prime}\overline{X}\, \gamma_\mu\, X^\prime\, G^\mu+{\cal M}_G^2 G_\mu G^{\mu\ast},
\ee
where $X$ and $X^\prime$ are any of the fermions $Q$, $q$, $U$, $D$, $t$, $b$, and all interaction terms allowed by the representation of the ETC boson under consideration should be taken into account. Integrating out the ETC boson $G_\mu$ at tree-level gives first
\be
G_{\mu}^\ast\sim-\frac{g_{XX^\prime}}{{\cal M}_G^2}{\overline X}\gamma_\mu X^\prime\ ,
\ee
and, after plugging back in ${\cal L}_{\rm{ETC}}^G$, one obtains the effective four fermion interaction
\be
{\cal L}_{\rm{ETC}}^G\sim-\frac{|g_{XX^\prime}|^2}{{\cal M}_G^2}\left|\overline{X}\, \gamma_\mu\, X^\prime\right|^2 \ ,
\ee
valid below the ETC scale. For example, for the class D and E bosons this procedure leads to
\be
{\cal L}_{\rm ETC}^{\rm D} = -\frac{|g_{Ub}|^2}{{\cal M}_D^2}\left(\overline{U}_R \gamma_\mu b_R\right)\left(\overline{b}_R \gamma^\mu U_R\right),\qquad
{\cal L}_{\rm ETC}^{\rm E}= -\frac{|g_{Dt}|^2}{{\cal M}_E^2}\left(\overline{D}_R \gamma_\mu t_R\right)\left(\overline{t}_R \gamma^\mu D_R\right)\ ,
\ee
where the quark color index $a=1,2,\dots N_c$ has been suppressed. The complete results of this classification are given in Appendix \ref{App:ETCclasses}. Also, it is convenient to Fierz rearrange some of the products of fermion bilinears. The formulas which we use are given in Appendix \ref{App:Fierz}.

Note that the diagonal couplings $g_{XX}$ are real, but the off-diagonal couplings $g_{XY}$, with $X\neq Y$, can be complex. We assume that also these couplings are real,
\begin{equation}
g_{XY}^\ast = g_{XY} \ ,
\end{equation}
{\em i.e.} we assume that there are no new sources of $CP$ violation.
Furthermore, for simplicity we assume that all ETC masses are identical:
\begin{equation}
{\cal M}_A={\cal M}_B={\cal M}_C={\cal M}_D={\cal M}_E \equiv {\cal M}\ .
\end{equation}
Putting together all ETC operators from Appendix \ref{App:ETCclasses} and using the Fierz rearrangement formulas from Appendix \ref{App:Fierz}, under the above assumptions, gives the ETC Lagrangian
\begin{eqnarray}
{\cal L}_{\rm ETC} &=& 2 G_{QqUt}\left[\left(\overline{Q}_L  U_R\right)\left(\overline{t}_R  q_L\right)+\left(\overline{q}_L  t_R\right)\left(\overline{U}_R  Q_L\right)\right]
+2G_{QqDb} \left[\left(\overline{Q}_L  D_R\right)\left(\overline{b}_R  q_L\right)+ \left(\overline{q}_L  b_R\right)\left(\overline{D}_R  Q_L\right)\right] \nonumber \\
&+&2G_{QQUU} \left(\overline{Q}_L  U_R\right)\left(\overline{U}_R Q_L\right)
+2G_{QQDD} \left(\overline{Q}_L  D_R\right)\left(\overline{D}_R Q_L\right)
+2G_{qqtt} \left(\overline{q}_L  t_R\right)\left(\overline{t}_R q_L\right)
+2G_{qqbb} \left(\overline{q}_L  b_R\right)\left(\overline{b}_R q_L\right) \nonumber \\
&+&\Delta {\cal L}_{\rm ETC} \ ,
\label{Eq:ETClagr}
\end{eqnarray}
where the couplings are defined as
\begin{eqnarray}
&& G_{QqUt}\equiv \frac{g_{Qq} g_{Ut}}{{\cal M}^2}\ , \quad
G_{QqDb}\equiv \frac{g_{Qq} g_{Db}}{{\cal M}^2}\ , \nonumber \\
&& G_{QQUU}\equiv \frac{g_{QQ} g_{UU}}{N {\cal M}^2}\ , \quad
G_{QQDD}\equiv \frac{g_{QQ} g_{DD}}{N {\cal M}^2}\ , \quad
G_{qqtt}\equiv \frac{g_{qq} g_{tt}}{N_c {\cal M}^2}\ , \quad
G_{qqbb}\equiv \frac{g_{qq} g_{bb}}{N_c {\cal M}^2}\ .
\label{Eq:MassOp}
\end{eqnarray}
The contribution $\Delta{\cal L}_{\rm{ETC}}$ is more complicated and reads
{\allowdisplaybreaks\allowdisplaybreaks[4]
\begin{align}\allowdisplaybreaks
\Delta {\cal L}_{\rm ETC} &=
 -\frac{1}{2}\frac{g_{QQ}^2}{{\cal M}^2}\left(\overline{Q}_L \gamma_\mu Q_L\right)^2
-\frac{1}{2}\frac{g_{qq}^2}{{\cal M}^2}\left(\overline{q}_L \gamma_\mu q_L\right)^2
-\frac{1}{2}\frac{g_{UU}^2}{{\cal M}^2}\left(\overline{U}_R \gamma_\mu U_R\right)^2
-\frac{1}{2}\frac{g_{DD}^2}{{\cal M}^2}\left(\overline{D}_R \gamma_\mu D_R\right)^2 \nonumber \\
&-\frac{1}{2}\frac{g_{tt}^2}{{\cal M}^2}\left(\overline{t}_R \gamma_\mu t_R\right)^2
-\frac{1}{2}\frac{g_{bb}^2}{{\cal M}^2}\left(\overline{b}_R \gamma_\mu b_R\right)^2
-\frac{g_{QQ}g_{qq}+g_{Qq}^2/2}{{\cal M}^2} \left(\overline{Q}_L \gamma_\mu Q_L\right)\left(\overline{q}_L \gamma^\mu q_L\right) \nonumber \\
&-\frac{g_{QQ}g_{tt}}{{\cal M}^2} \left(\overline{Q}_L \gamma_\mu Q_L\right)\left(\overline{t}_R \gamma^\mu t_R\right)
-\frac{g_{QQ}g_{bb}}{{\cal M}^2} \left(\overline{Q}_L \gamma_\mu Q_L\right)\left(\overline{b}_R \gamma^\mu b_R\right)
-\frac{g_{qq}g_{UU}}{{\cal M}^2} \left(\overline{q}_L \gamma_\mu q_L\right)\left(\overline{U}_R \gamma^\mu U_R\right) \nonumber \\
&-\frac{g_{qq}g_{DD}}{{\cal M}^2} \left(\overline{q}_L \gamma_\mu q_L\right)\left(\overline{D}_R \gamma^\mu D_R\right)
-\frac{g_{UU}g_{DD}}{{\cal M}^2} \left(\overline{U}_R \gamma_\mu U_R\right)\left(\overline{D}_R \gamma^\mu D_R\right)
-\frac{g_{UU}g_{tt}+g_{Ut}^2}{{\cal M}^2} \left(\overline{U}_R \gamma_\mu U_R\right)\left(\overline{t}_R \gamma^\mu t_R\right) \nonumber \\
&- \frac{g_{UU}g_{bb}+g_{Ub}^2}{{\cal M}^2} \left(\overline{U}_R \gamma_\mu U_R\right)\left(\overline{b}_R \gamma^\mu b_R\right)
-\frac{g_{DD}g_{tt}+g_{Dt}^2}{{\cal M}^2} \left(\overline{D}_R \gamma_\mu D_R\right)\left(\overline{t}_R \gamma^\mu t_R\right) \nonumber \\
&- \frac{g_{DD}g_{bb}+g_{Db}^2}{{\cal M}^2} \left(\overline{D}_R \gamma_\mu D_R\right)\left(\overline{b}_R \gamma^\mu b_R\right)
-\frac{g_{tt}g_{bb}}{{\cal M}^2} \left(\overline{t}_R \gamma_\mu t_R\right) \left(\overline{b}_R \gamma^\mu b_R\right)
-\frac{g_{UD}^2}{{\cal M}^2} \left(\overline{U}_R \gamma_\mu D_R\right) \left(\overline{D}_R \gamma^\mu U_R\right) \nonumber \\
&-\frac{g_{tb}^2}{{\cal M}^2} \left(\overline{t}_R \gamma_\mu b_R\right) \left(\overline{b}_R \gamma^\mu t_R\right)
-\frac{g_{UD} g_{tb}+g_{Ut} g_{Db}}{{\cal M}^2}\left[\left(\overline{U}_R \gamma_\mu D_R\right)  \left(\overline{b}_R \gamma^\mu t_R\right)
+\left(\overline{D}_R \gamma_\mu U_R\right)\left(\overline{t}_R \gamma^\mu b_R\right) \right] \nonumber \\
&-\frac{2g_{Qq}^2}{{\cal M}^2}\left(\overline{Q}_L \gamma_\mu T^i Q_L\right)\left(\overline{q}_L \gamma^\mu T^i q_L\right)
+\frac{4g_{QQ}g_{UU}}{{\cal M}^2} \left(\overline{Q}_L T_{\rm TC}^A U_R\right)\left(\overline{U}_R T_{\rm TC}^A Q_L\right) \nonumber \\
&+\frac{4g_{QQ}g_{DD}}{{\cal M}^2} \left(\overline{Q}_L T_{\rm TC}^A D_R\right)\left(\overline{D}_R T_{\rm TC}^A Q_L\right)
+\frac{4g_{qq}g_{tt}}{{\cal M}^2} \left(\overline{q}_L T_{\rm QCD}^a t_R\right)\left(\overline{t}_R T_{\rm QCD}^a q_L\right) \nonumber \\
&+\frac{4g_{qq}g_{bb}}{{\cal M}^2} \left(\overline{q}_L T_{\rm QCD}^a b_R\right)\left(\overline{b}_R T_{\rm QCD}^a q_L\right) \ ,
\label{Eq:DETC}
\end{align}}
where $T_{\rm TC}^A$ are the TC generators for the $N$ representation, and $T_{\rm QCD}^a$ are the generators for the fundamental representation of $SU(N_c)$. These matrices are normalized as
\begin{equation}
{\rm Tr}\ T^i T^j = \frac{1}{2}\delta^{ij}\ , \quad
{\rm Tr}\ T_{\rm QCD}^a T_{\rm QCD}^b = \frac{1}{2}\delta^{ab}\ , \quad
{\rm Tr}\ T_{\rm TC}^A T_{\rm TC}^B = \frac{1}{2}\delta^{AB}\ .
\end{equation}
As we shall see below, the operators not included in $\Delta {\cal L}_{\rm ETC}$ contribute both to scalar and fermion masses, whereas the operators included in $\Delta {\cal L}_{\rm ETC}$ contribute to neither.\footnote{Note that in TC theories with near-conformal dynamics, four-fermion operators with techniquark bilinears may be enhanced relative to operators with quark bilinears. In this paper we will not pursue such more model dependent questions, but treat all four fermion interactions appearing in (\ref{Eq:ETClagr}).}

We compute observables in the large-$N$ limit, with $N/N_c$ finite. For a consistent large-$N$ expansion, the ETC couplings $g_{XY}$ must scale like $1/\sqrt{N}$. Therefore, the $G_{QqUt}$ and $G_{QqDb}$ couplings in (\ref{Eq:MassOp}) scale like $1/N$, whereas the diagonal couplings $G_{QQXX}$ and $G_{qqXX}$ scale like $1/N^2$, the extra factor of $1/N$ arising from the Fierz rearrangement of the class-A operators. It is clear that the operators in (\ref{Eq:MassOp}) contribute to mass, as they involve bilinears mixing left-handed and right-handed fermions. Therefore, we get mass contribution at leading order (LO) in $N$ from the $G_{QqUt}$ and $G_{QqDb}$ operators, and contributions at next-to-leading order (NLO) from the $G_{QQXX}$ and $G_{qqXX}$ operators. On the other hand, the operators contained in $\Delta {\cal L}_{\rm ETC}$ contribute to fermion and scalar masses neither to LO nor to NLO in the large-$N$ expansion. At LO this is evident from the presence of uncontracted $\gamma_\mu$ matrices in separate loops. The NLO is zero either because of the appearance of products $P_L P_R=0$ (as in the case of left-left bilinear products and the operators with the TC and QCD generators), or because it is manifestly absent (as in the case of left-right bilinear products mixing quarks and techniquarks). Hence, we can consistently compute masses to LO and NLO by only considering the operators in (\ref{Eq:MassOp}). However, NLO computations are rather complicated. In this paper, we find it more convenient to formally treat the $G_{QQXX}$ and $G_{qqXX}$ couplings as quantities scaling like $1/N$, and compute all observables to LO in the large-$N$ expansion. The error is still NLO in $1/N$, but this approach allows us to account for the important mass contribution from the class-A operators\footnote{This approach is similar to the one adopted in topcolor-assisted technicolor for treating the new hypercharge interactions.}.
\section{Cutoff and confinement}\label{Sec:Cutoff}
There are two physical cutoffs in the model: $\Lambda$, associated to TC dynamics, and the ETC scale ${\cal M}$. Therefore, we are naturally led to use a cutoff regulator for the standard loop integrals. It is not clear, though, which one of the two cutoffs should be used to evaluate the integrals. A possible approach consists in using the smaller mass scale, which we assume to be $\Lambda$. This, however, would imply losing information from the dynamics occurring between $\Lambda$ and ${\cal M}$. Furthermore, it is well know that making the techniquark loop integrals finite with a sharp cutoff does not account for confinement, as the fermion propagators go on-shell for sufficiently large external momenta.  A solution to both problems is provided by models of confinement. In the model of \cite{Efimov:1993zg}, for instance, the interaction of $n$ external mesons is given by amplitudes of the form
\begin{eqnarray}
&&i\, T(q_1,q_2,\dots,q_{n-1})\equiv 
-\int\frac{d^4k}{(2\pi)^4}\, \int\frac{dM_Q}{2\pi i}\, \rho(M_Q)\, {\rm Tr}\, 
i\, \Gamma_1\, \frac{i(\slashed{k}-\slashed{q}_1+M_Q)}{(k-q_1)^2-M_Q^2}\, \nonumber \\
&&i\, \Gamma_2\, \frac{i(\slashed{k}-\slashed{q}_1-\slashed{q}_2+M_Q)}{(k-q_1-q_2)^2-M_Q^2}\cdots\, 
i\, \Gamma_{n-1}\, \frac{i(\slashed{k}-\slashed{q}_1-\slashed{q}_2-\dots -\slashed{q}_{n-1}+M_Q)}{(k-q_1-q_2-\dots -q_{n-1})^2-M_Q^2}\,
i\, \Gamma_n\, \frac{i(\slashed{k}+M_Q)}{k^2-M_Q^2}\ , \nonumber \\
&\ & 
\end{eqnarray}
where $\Gamma_i$ are matrices in Dirac space, and are determined by the quantum numbers of the external mesons. Here the fermion mass $M_Q$ is a complex variable which is integrated along a closed contour enclosing the external momenta. Confinement and convergence of the integrals are both guaranteed by taking the function $\rho(z)$ to be holomorphic everywhere and decreasing faster than any polynomial for $|z|\to\infty$. Under these assumptions we may use the Cauchy integral formula to obtain
\begin{eqnarray}
\rho(z)=\int\frac{dM_Q}{2\pi i}\, \frac{\rho(M_Q)}{M_Q-z} = \frac{1}{\kappa}\, a(-z^2/\kappa^2)+\frac{1}{\kappa^2}\, z\, b(-z^2/\kappa^2)\ ,
\end{eqnarray}
where $\kappa$ is an intrinsic mass scale of confinement, and $a(\xi), b(\xi)\to 0$ faster than any polynomial for $|\xi|\to\infty$. The equation above gives
\begin{eqnarray}
a(-z^2/\kappa^2) = \kappa\, \int\frac{dM_Q}{2\pi i}\, \frac{\rho(M_Q)\, M_Q}{M_Q^2-z^2} \ , \quad
b(-z^2/\kappa^2) = \kappa^2\, \int\frac{dM_Q}{2\pi i}\, \frac{\rho(M_Q)}{M_Q^2-z^2} \ .
\end{eqnarray}
Consider for instance the two-point function for two external scalar mesons, that is $\Gamma_1=\Gamma_2=1$:
\begin{eqnarray}
&&i\, T(q)\equiv 
-\int\frac{d^4k}{(2\pi)^4}\, \int\frac{dM_Q}{2\pi i}\, \rho(M_Q)\, {\rm Tr}\, 
i\,  \frac{i(\slashed{k}-\slashed{q}+M_Q)}{(k-q_1)^2-M_Q^2}\, 
i\,  \frac{i(\slashed{k}+M_Q)}{k^2-M_Q^2}\ ,  
\end{eqnarray}
After combining the denominators, reducing the powers of $M_Q$ in the numerator, Wick rotating, shifting to Euclidean momentum, and changing the integration variable to $u\equiv k_E^2$, we obtain
\begin{eqnarray}
&&i\, T(q) = \frac{i}{4\pi^2}\frac{1}{\kappa^2}\int_0^1 dx \int_0^\infty du\left[-2u^2\frac{d}{du}-u\right]b\left(u/\kappa^2-x(1-x)q^2/\kappa^2\right) \ .
\end{eqnarray}
Integrating by parts, and using the hypothesis that $b(\xi)$ decreases faster than any polynomial for $|\xi|\to\infty$, leads to the result
\begin{eqnarray}
T(q) = \frac{3}{4\pi^2}\left[\kappa^2\, B_1(q^2)+q^2\, B_0(q^2)\right]\ ,
\label{Eq:T2}
\end{eqnarray}
where
\begin{eqnarray}
B_0(q^2) &\equiv& \int_0^1 dx\, \int_0^\infty d\xi \, x(1-x)\, b(\xi-x(1-x)q^2/\kappa^2)\ , \nonumber \\
B_1(q^2) &\equiv& \int_0^1 dx\, \int_0^\infty d\xi \, \left(\xi-x(1-x)q^2/\kappa^2\right)\, b(\xi-x(1-x)q^2/\kappa^2)\ .
\end{eqnarray}
These functions are finite and, featuring no pole singularity, imply fermion confinement. 

It is interesting to compute the three-meson and four-meson interactions at zero external momenta. For three external mesons we have to compute integrals like
\begin{eqnarray}
-\int\frac{d^4k}{(2\pi)^4}\, \int\frac{dM_Q}{2\pi i}\, \rho(M_Q)\, {\rm Tr}\, 
i\, \Gamma_1\, \frac{i(\slashed{k}+M_Q)}{k^2-M_Q^2}\, 
i\, \Gamma_2\, \frac{i(\slashed{k}+M_Q)}{k^2-M_Q^2}\, 
i\, \Gamma_3\, \frac{i(\slashed{k}+M_Q)}{k^2-M_Q^2}\, \nonumber
\end{eqnarray}
In general, this requires evaluating an integral of the form
\begin{eqnarray}
i\, I(c_1,c_3)\equiv \int\frac{d^4k}{(2\pi)^4}\, \int\frac{dM_Q}{2\pi i}\, \rho(M_Q)\, \frac{c_1\, M_Q\, k^2+c_3\, M_Q^3}{(k^2-M_Q^2)^3}\ ,
\end{eqnarray}
where the coefficients $c_1$ and $c_3$ depend on the $\Gamma_i$ matrices. Employing the same techniques leading to (\ref{Eq:T2}) gives
\begin{eqnarray}
I(c_1,c_3) = \frac{1}{16\pi^2}\, \frac{1}{\kappa}\int_0^\infty du\, \left[(c_1+c_3)\frac{u^2}{2}\frac{d^2}{du^2}+c_3\, u\, \frac{d}{du}\right]a(u/\kappa^2)\ .
\end{eqnarray}
Integrating by parts twice, and using the hypothesis that $a(\xi)$ decreases faster than any polynomial for $|\xi|\to\infty$, leads to
\begin{eqnarray}
I(c_1,c_3) = \frac{c_1\, \kappa}{16\pi^2}\int_0^\infty d\xi\, a(\xi)\ .
\label{Eq:3point}
\end{eqnarray}
In the case of four external mesons at zero external momenta, the integrals to be computed are like
\begin{eqnarray}
-\int\frac{d^4k}{(2\pi)^4}\, \int\frac{dM_Q}{2\pi i}\, \rho(M_Q)\, {\rm Tr}\, 
i\, \Gamma_1\, \frac{i(\slashed{k}+M_Q)}{k^2-M_Q^2}\, 
i\, \Gamma_2\, \frac{i(\slashed{k}+M_Q)}{k^2-M_Q^2}\, 
i\, \Gamma_3\, \frac{i(\slashed{k}+M_Q)}{k^2-M_Q^2}\,
i\, \Gamma_4\, \frac{i(\slashed{k}+M_Q)}{k^2-M_Q^2}\ . \nonumber 
\end{eqnarray}
This requires evaluating an integral of the form
\begin{eqnarray}
i\, I(c_0,c_2,c_4)\equiv -\int\frac{d^4k}{(2\pi)^4}\, \int\frac{dM_Q}{2\pi i}\, \rho(M_Q)\, \frac{c_0\, (k^2)^2+c_2\, M_Q^2\, k^2+c_4\, M_Q^4}{(k^2-M_Q^2)^4}\ ,
\end{eqnarray}
which eventually gives
\begin{eqnarray}
I(c_0,c_2,c_4) = \frac{c_0}{16\pi^2}\int_0^\infty d\xi\, b(\xi)\ .
\label{Eq:4point}
\end{eqnarray}
The interesting aspect of (\ref{Eq:3point}) and (\ref{Eq:4point}) is that only the highest power of momentum contributes to the loop integral. This is important, as it preserves the special relation between form factors which is implied by the underlying chiral symmetry. In fact, using a sharp cutoff, rather than a confining function, the terms with the highest power of loop momentum, in the three-point and four-point vertices, correspond to the leading divergent logarithm, which preserves the underlying chiral symmetry \cite{Delbourgo:1993dk}. 

If we use a distribution density $\rho(M)$ to smear the integrals over techniquarks, we may cutoff the full theory at ${\cal M}$.  The integrals over SM quarks are cutoff at ${\cal M}$, whereas the integrals over techniquarks are naturally finite. Clearly we must choose an appropriate function $\rho(M)$, and integrals are unavoidably more difficult to evaluate than the standard loop integrals, especially in the presence of isospin mass splitting. However in our analysis we are only interested in small external momenta, and thus we are not concerned with unphysical thresholds. Therefore, we make the approximation of using a sharp cutoff  $\Lambda$ for the loop integrals over techniquark momenta, rather than a distribution density, while still cutting off the SM-fermion loop integrals at ${\cal M}$. This approach allows the dynamics between $\Lambda$ and ${\cal M}$ to contribute to the low-energy observables, and 
at the same time preserves the symmetries of the Lagrangian \cite{DiChiara:2014gsa}. In accordance with the above results, our prescription is the following:
\begin{enumerate}
\item Compute integrals over techniquarks with a cutoff $\Lambda$, and integrals over ordinary quarks with a cutoff ${\cal M}$.
\item In evaluating interaction vertices, retain only the logarithmically divergent part of the integral.
\item Evaluate the integrals at zero external momenta. 
\end{enumerate}

We will need to evaluate fermion loops with external weak bosons, hence we must use a regulator preserving gauge invariance. Since we are using a cutoff, we find it convenient to employ the regularization prescription of \cite{Cynolter:2010ei}, and require that the relation
\begin{eqnarray}
\int\, \frac{d^4l_E}{(2\pi)^4}\, \frac{l_{E\mu}l_{E\nu}}{(l_E^2+m^2)^{n+1}} = \frac{g_{\mu\nu}}{2n} \int \frac{d^4l_E}{(2\pi)^4} \frac{1}{(l_E^2+m^2)^n}
\end{eqnarray}
is satisfied, for integrals in Euclidean space, for any $n\geq 1$. After this condition is imposed, integrals may be evaluated with a sharp cutoff. In \cite{Cynolter:2010ei} this prescription is shown to satisfy the Ward identities.

We end this section by enlisting the standard integrals used for computing the two-point functions. In accordance to the prescription above, we evaluate these at zero external momentum:
\begin{eqnarray}
I_X &\equiv&   i\int\frac{d^4 k}{(2\pi)^4}\frac{1}{k^2-M_X^2}\ , \nonumber \\
J_{XY} &\equiv&  -i\int_0^1 dx\int\frac{d^4 l}{(2\pi)^4}\frac{1}{\left(l^2-x\, M_X^2-(1-x)M_Y^2\right)^2} \ , \nonumber \\
K_{XY} &\equiv&  -i\int_0^1 dx\int\frac{d^4 l}{(2\pi)^4}\frac{x}{\left(l^2-x\, M_X^2-(1-x)M_Y^2\right)^2} \ , \nonumber \\
L_{XY} &\equiv&  -i\int_0^1 dx\int\frac{d^4 l}{(2\pi)^4}\frac{x(1-x)}{\left(l^2-x\, M_X^2-(1-x)M_Y^2\right)^2} \ .
\end{eqnarray}
In order to evaluate the scalar wavefunction renormalization, as well as the $S$ parameter, we also need to consider $J_{XX}$ at finite external momentum $q$, take the derivative with respect to $q^2$, and evaluate the resulting integral at $q^2=0$. Since the latter has dimension $1/M_X^{2}$, we find it convenient to define
\begin{eqnarray}
J_{XX}^\prime &\equiv&  \frac{i}{3}\int\frac{d^4 l}{(2\pi)^4}\frac{M_X^2}{\left(l^2-M_X^2\right)^3} \ .
\end{eqnarray}
We provide explicit expressions for these integrals in Appendix \ref{Sec:Int}.
\section{Mass of the fundamental particles}\label{Sec:Masses}
\subsection{Fermion masses}
\begin{figure}[t!]
\includegraphics[width=4.5in]{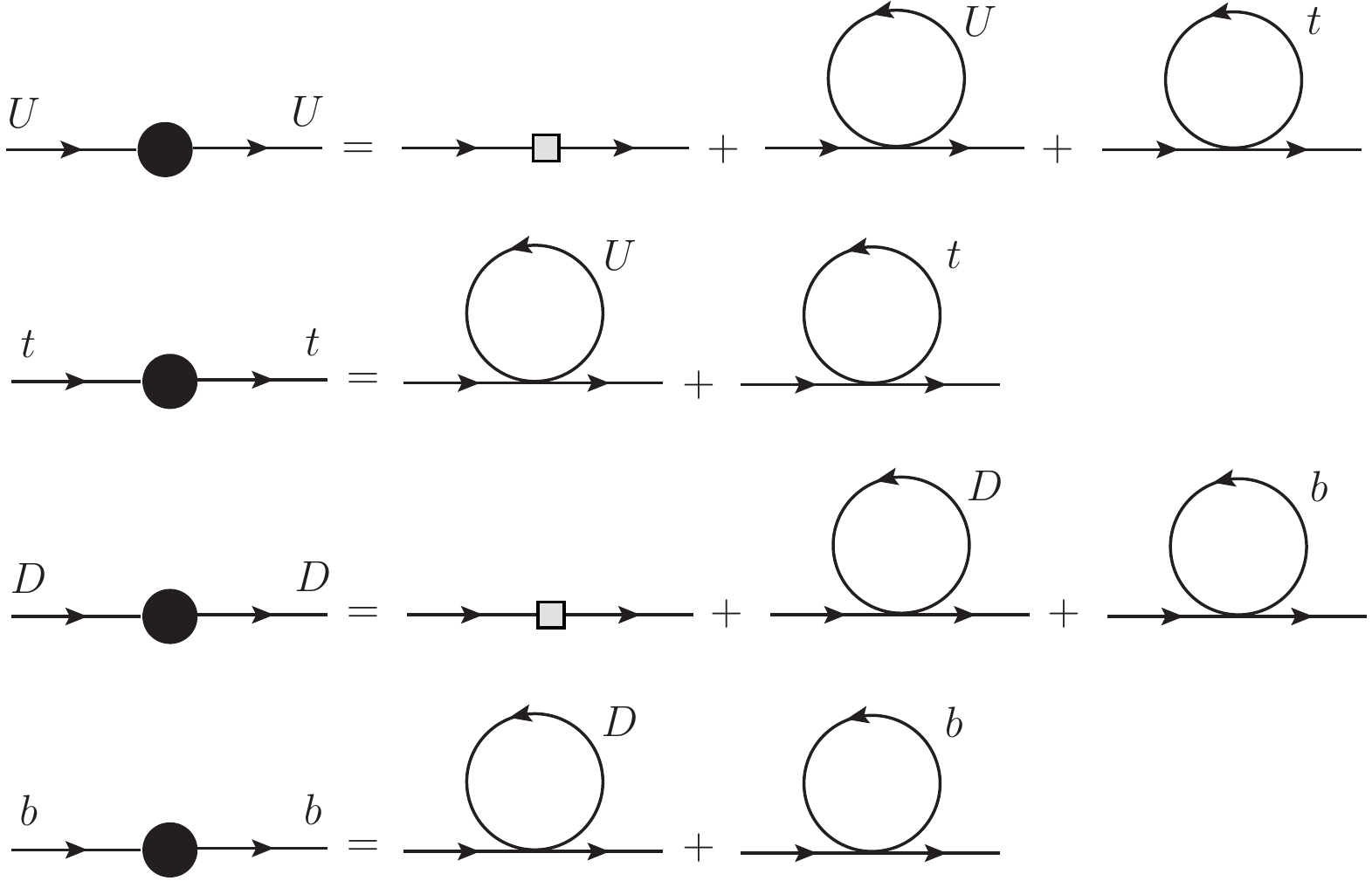}
\caption{Diagrams contributing to the fermion masses at LO in the large-$N$ expansion. The dominant contribution to $U$ and $D$ mass arise from TC dynamics, and is denoted by the tree-level mass insertions.}
\label{Fig:Gap}
\end{figure}
To LO in the large-$N$ expansion, the fermion masses are given by the diagrams of Fig.~\ref{Fig:Gap}.  These lead to the coupled equations
\begin{eqnarray}
M_U &=& M_Q + 4\, N\, G_{QQUU}\, M_U\, I_U + 4\, N_c\, G_{QqUt}\, M_t\, I_t \nonumber \\
M_t &=& 4\, N\, G_{QqUt}\, M_U\, I_U + 4\, N_c\, G_{qqtt}\, M_t\, I_t \ , 
\label{Eq:Ut}
\end{eqnarray}
and
\begin{eqnarray}
M_D &=& M_Q + 4\, N\, G_{QQDD}\, M_D\, I_D + 4\, N_c\, G_{QqDb}\, M_b\, I_b \nonumber \\
M_b &=& 4\, N\, G_{QqDb}\, M_D\, I_D + 4\, N_c\, G_{qqbb}\, M_b\, I_b \ .
\label{Eq:Db}
\end{eqnarray}
Note that unlike the model of \cite{DiChiara:2014gsa}, in which the only ETC operator was the one proportional to $G_{QqUt}$, now we have additional ETC contributions to mass. In particular, the $U-D$ isospin splitting may be softened or even set to zero by adjusting the $G_{QQUU}$ and $G_{QQDD}$ operators. This removes the major obstacle of \cite{DiChiara:2014gsa}: There, in order to obtain a large reduction of the TC-Higgs mass, the value of $G_{QqUt}\times{\cal M}$ had to be increased. This, in turn, made $U$ considerably heavier than $D$, and the $T$ parameter unacceptably large. Now, instead, contributions from $G_{QQUU}$ and $G_{QQDD}$ have a double effect: they reduce the amount of $U-D$ isospin splitting, and, as we shall see, contribute to further reduce the TC-Higgs mass.
\subsection{Weak boson masses}
\begin{figure}[t!]
\includegraphics[width=4.5in]{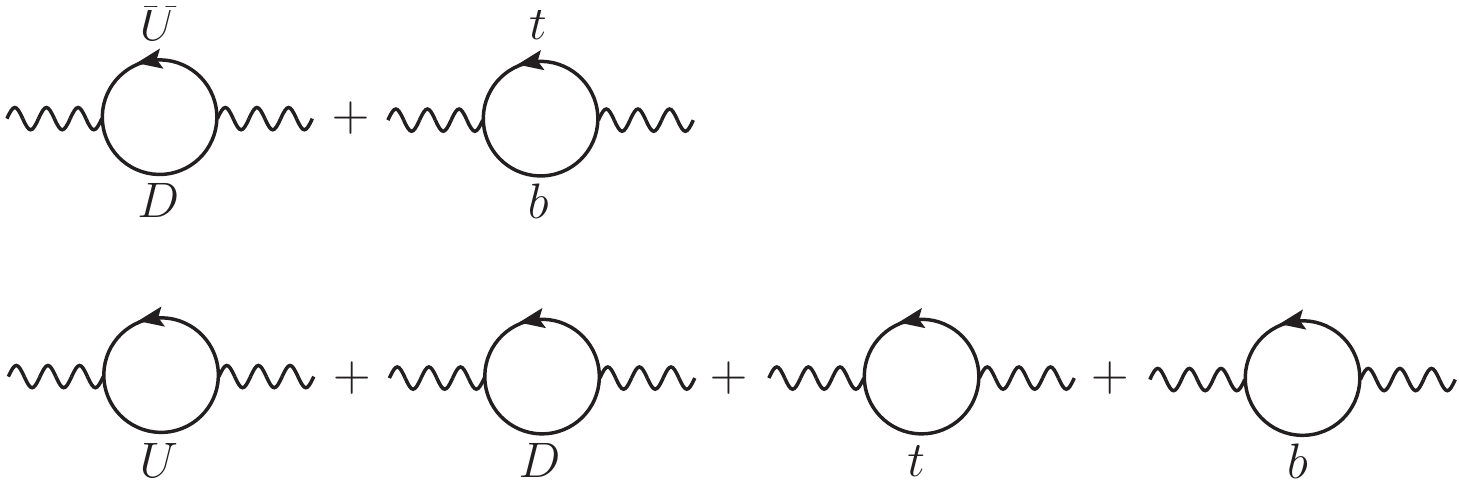}
\caption{Diagrams contributing to the coefficient of $g^{\mu\nu}$ in the $W$ (top) and $Z$ boson (bottom) VPAs. Contributions from the operators contained in $\Delta{\cal L}_{\rm ETC}$ are suppressed by a factor of the order of $M_Q^2/{\cal M}^2$, and may therefore be ignored.}
\label{Fig:MW}
\end{figure}
We may compute the $W$ mass in terms of the fermion masses. In order to do so we must compute the corresponding vacuum-polarisation amplitude (VPA), which is required by gauge invariance to be transverse:
\begin{equation}
\Pi_{WW}^{\mu\nu}(q) = \Pi_{WW}(q^2)\left(g^{\mu\nu}-\frac{q^\mu q^\nu}{q^2}\right)\ .
\end{equation}
The expression for $\Pi_{WW}(q^2)$ can be extracted from the $g^{\mu\nu}$ part of the amplitude. Ignoring contributions from $\Delta{\cal L}_{\rm ETC}$, which are suppressed by a factor of the order of $M_Q^2/{\cal M}^2$ \footnote{The $\Delta{\cal L}_{\rm ETC}$ contribution to $M_W$ and $M_Z$ may be ignored as long as the corresponding contribution to the $T$ parameter is within experimental bounds, which is anyway a strict requirement for viability.}, the only contribution to $g^{\mu\nu}$ arises from the one-loop diagrams of Fig.~\ref{Fig:MW} (top). In order to recover a fully transverse result, one needs to include an infinite chain of fermion loops, as well as tree-level Goldstone bosons exchanges. Using the fermion mass equations, we have verified in the simplified case $G_{QQUU}, G_{QQDD}\to 0$ that transversality is recovered. From the one-loop diagrams we obtain
\begin{eqnarray}
\Pi_{WW}(q^2) &=& -2 g^2 \Big[N\, L_{UD}+N_c\, L_{tb}\Big]q^2 + g^2\Big[N\, M_U^2\, K_{UD}+N\, M_D^2\, K_{DU} + N_c\, M_t^2\, K_{tb} + N_c\, M_b^2\, K_{bt}\Big]\ .
\end{eqnarray}
Since the $W$ boson has a tree-level kinetic term, to leading-order in the weak coupling $g$ we may ignore the first term. Then $M_W$ is given by \footnote{Clearly the contribution from $M_b$ is completely negligible: however we display it in order to show the contribution from all isospin components.}
\begin{equation}
M_W^2 = g^2\left[N\, M_U^2\, K_{UD}+N\, M_D^2\, K_{DU} + N_c\, M_t^2\, K_{tb} + N_c\, M_b^2\, K_{bt}\right]\ .
\label{Eq:MW}
\end{equation}
We may rewrite this equation as
\begin{equation}
\frac{1}{\sqrt2\, G_F} = 4\left[N\, M_U^2\, K_{UD}+N\, M_D^2\, K_{DU} + N_c\, M_t^2\, K_{tb} + N_c\, M_b^2\, K_{bt}\right]\ ,
\label{Eq:v}
\end{equation}
where $G_F$ is the Fermi constant, $(\sqrt2\, G_F)^{-1}\simeq 246$ GeV. Equation (\ref{Eq:v}) generalizes the Pagels-Stokar equation by taking into account the ETC contributions. Using this equation, together with the fermion mass equations (\ref{Eq:Ut}) and (\ref{Eq:Db}), we can solve for $M_Q$, $M_U$, $M_D$, $G_{QqUt}$, $G_{QqDb}$, as a function of $\Lambda$, ${\cal M}$, $G_{QQUU}$, $G_{QQDD}$, $G_{qqtt}$, $G_{qqbb}$, $N$, and the experimental values of $G_F$, $M_t$, $M_b$ and $N_c$. 

We finally compute the mass of the $Z$ boson. The result is
\begin{eqnarray}
M_Z^2 = \frac{g^2+g^{\prime 2}}{2}\left[N\, M_U^2\, J_{UU}+N\, M_D^2\, J_{DD}+N_c\, M_t^2\, J_{tt}+N_c\, M_b^2\, J_{bb}\right]\ .
\end{eqnarray}
\section{Mass of the TC Higgs}\label{Sec:scalarmass}
\begin{figure}[t!]
\includegraphics[width=5.0in]{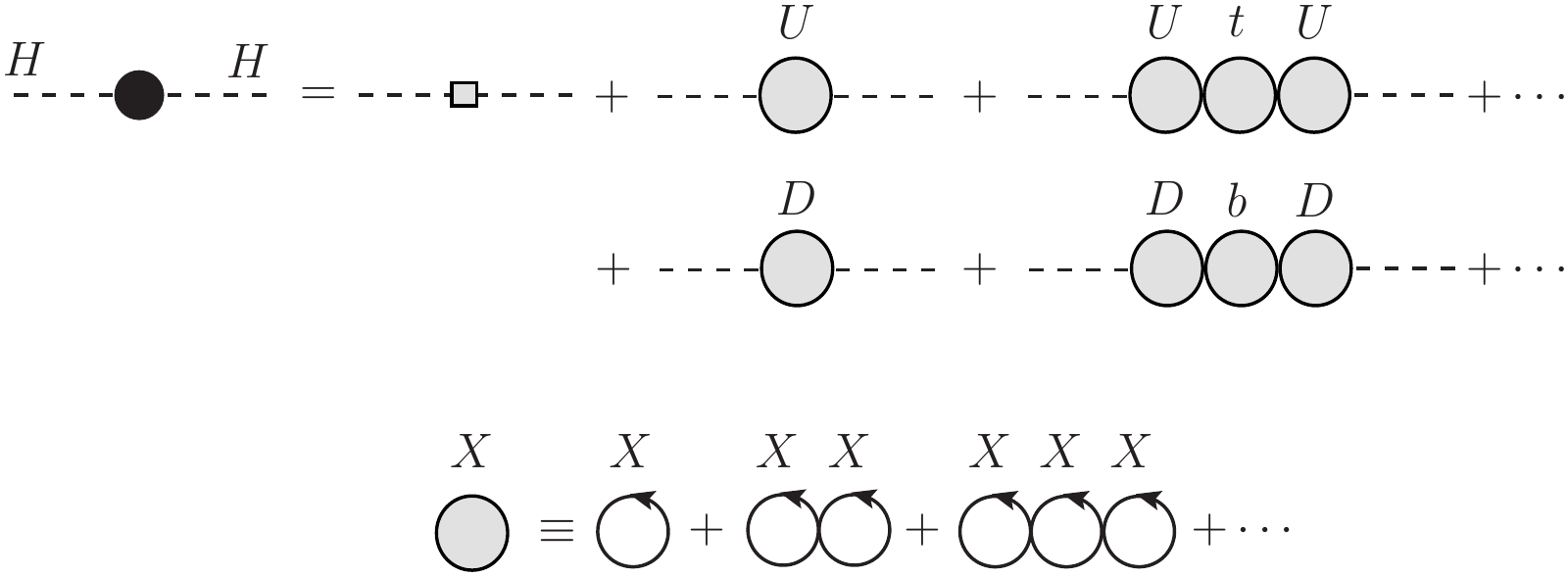}
\caption{The diagrams contributing to the TC-Higgs self-energy are shown in the top figure, including the tree-level mass. The shaded circles represent the sum of fermion-bubble chains, as shown in the bottom figure.}
\label{Fig:MH}
\end{figure}

The TC-Higgs self-energy is given by the chain of diagrams shown in Fig.~\ref{Fig:MH}. Including the tree-level mass $M^2$, we obtain
\begin{eqnarray}
\Sigma_{HH} = -M^2 + N y^2 \Bigg[\frac{{\cal I}_{UU}^{SS}}{1-N\, N_c\, G_{QqUt}^2\, {\cal I}_{UU}^{SS}\, {\cal I}_{tt}^{SS}} 
+ \frac{{\cal I}_{DD}^{SS}}{1-N\, N_c\, G_{QqDb}^2\, {\cal I}_{DD}^{SS}\, {\cal I}_{bb}^{SS}}\Bigg]\ ,
\label{Eq:HH}
\end{eqnarray}
where
\begin{equation}
{\cal I}_{XX}^{SS} \equiv \left\{
\begin{array}{lr}
\vspace{0.3cm}
\displaystyle{\frac{I_{XX}^{SS}}{1-N\, G_{QQXX}\, I_{XX}^{SS}}} ,& X=U,D \\
\displaystyle{\frac{I_{XX}^{SS}}{1-N_c\, G_{qqXX}\, I_{XX}^{SS}}}, & X=t,b\ ,
\end{array}
\right.
\label{Eq:ISS}
\end{equation}
and
\begin{eqnarray}
I_{XY}^{SS} &\equiv&-i \int \frac{d^4 k}{(2\pi)^4}\, {\rm Tr}\, \frac{i\left(\slashed{k}+M_Y\right)}{k^2-M_Y^2+i\, \varepsilon} \frac{i\left(\slashed{k}+\slashed{q}+M_X\right)}{(k+q)^2-M_X^2+i\, \varepsilon}
=2\left(q^2-(M_X+M_Y)^2\right)J_{XY}+2(I_X+I_Y)\ .
\label{Eq:ISSintegral}
\end{eqnarray}
We trade $M$ for the one-loop mass $M_{H0}$:
\begin{equation}
-M^2+N y^2 \left(I_{UU}^{SS}+I_{DD}^{SS}\right)_{q^2=M_{H0}^2}=0\ .
\label{Eq:MH0}
\end{equation}
This is a quantity with a precise physical meaning: after including the ETC corrections to the TC vacuum, which are encoded by the expression for $G_F$ in  (\ref{Eq:v}), $M_{H0}$ is the scalar mass due to TC dynamics alone, and can be estimated, for instance, by scaling up the mass of the $\sigma$ meson from QCD. Using this definition, $\Sigma_{HH}$ becomes
\begin{eqnarray}
\Sigma_{HH} &=& N y^2 \Bigg[
\frac{{\cal I}_{UU}^{SS} }{1-N\, N_c\, G_{QqUt}^2\, {\cal I}_{UU}^{SS} \, {\cal I}_{tt}^{SS} } 
+ \frac{{\cal I}_{DD}^{SS} }{1-N\, N_c\, G_{QqDb}^2\, {\cal I}_{DD}^{SS} \, {\cal I}_{bb}^{SS} }
-\left(I_{UU}^{SS}+I_{DD}^{SS}\right)_{q^2=M_{H0}^2}\Bigg]\ .
\label{Eq:HHself}
\end{eqnarray}
The physical Higgs mass, $M_H\simeq 125$ GeV, is found by solving the equation
\begin{equation}
\Sigma_{HH}\left(M_H^2\right) = 0\ . 
\end{equation}
This can be solved for $M_{H0}$ by using (\ref{Eq:ISSintegral}) in (\ref{Eq:HHself})
\begin{eqnarray}
M_{H0}^2 &=& \frac{1}{2\left(J_{UU}+J_{DD}\right)}\Bigg[\Bigg(
\frac{{\cal I}_{UU}^{SS}}{1-N\, N_c\, G_{QqUt}^2\, {\cal I}_{UU}^{SS}\, {\cal I}_{tt}^{SS}} 
+ \frac{{\cal I}_{DD}^{SS}}{1-N\, N_c\, G_{QqDb}^2\, {\cal I}_{DD}^{SS}\, {\cal I}_{bb}^{SS}} \Bigg)_{q^2=M_H^2} \nonumber \\
&&\quad\quad\quad\quad\quad\ \ \  +8\left(M_U^2\, J_{UU}+M_D^2\, J_{DD}\right)
-4 \left(I_U+I_D\right)\Bigg]\ .
\label{Eq:CondMH0}
\end{eqnarray}
To understand the implications of (\ref{Eq:CondMH0}), we assume for a moment that we can expand in the ETC couplings. Expanding up to the linear term in the four-fermion couplings, and ignoring corrections of the order of $M_X^2/\Lambda^2$, leads to the approximate relation
\begin{eqnarray}
M_H^2 &\simeq& M_{H0}^2 - \frac{N\, \left(G_{QQUU}+G_{QQDD}\right)\Lambda^2}{4\pi^2}\,  \frac{\Lambda^2}{\langle {\overline \log}(\Lambda^2/M_Q^2)\rangle}
\label{Eq:Estimate}
\end{eqnarray}
where we defined
\begin{equation}
{\overline \log}\frac{\Lambda^2}{M_X^2}\equiv \log\frac{\Lambda^2}{M_X^2}-1\ .
\end{equation}
and
\begin{eqnarray}
\langle {\overline \log}\frac{\Lambda^2}{M_Q^2}\rangle\equiv \frac{1}{2} \left({\overline \log}\, \frac{\Lambda^2}{M_U^2}+{\overline \log}\, \frac{\Lambda^2}{M_D^2}\right)
\end{eqnarray}

The last factor in (\ref{Eq:Estimate}) is reasonably of the order of (1 TeV)$^2$. Estimates obtained by scaling up the mass of the $\sigma$ meson from QCD to a TC theory suggest $M_{H0}\lesssim 1$ TeV \cite{Foadi:2012bb}. Therefore, the factor containing the ETC couplings, in (\ref{Eq:Estimate}), must be
${\cal O}(1)$. Bounds on the ETC couplings can be derived by requiring that the series leading to (\ref{Eq:HH}) and (\ref{Eq:ISS}) are convergent. This gives the constraints
\begin{eqnarray}
N\, G_{QQUU}\, \Lambda^2 < 4\pi^2\ , \quad
N\, G_{QQDD}\, \Lambda^2 < 4\pi^2\ , \quad
N_c\, G_{qqtt}\, {\cal M}^2 < 4\pi^2\ , \quad
N_c\, G_{qqbb}\, {\cal M}^2 < 4\pi^2 \ ,
\label{Eq:Constr1}
\end{eqnarray}
which guarantee, respectively, no $\overline{U}U$, $\overline{D}D$, $\overline{t}t$, and $\overline{b}b$ condensation from ETC, and the constraints
\begin{eqnarray}
N\, N_c\, G_{QqUt}^2\, \Lambda^2\, {\cal M}^2< 16\pi^4\ , \quad
N\, N_c\, G_{QqDb}^2\, \Lambda^2\, {\cal M}^2< 16\pi^4\ ,
\label{Eq:Constr2}
\end{eqnarray}
which prevent the ETC force from generating $\overline{U}t$ and $\overline{D}b$ condensates, respectively. If the ETC couplings are large enough, yet subcritical, we obtain $N\, \left(G_{QQUU}+G_{QQDD}\right)\Lambda^2/4\pi^2\sim 1$, which, according to the estimate (\ref{Eq:Estimate}), is the order of magnitude required for the ETC interactions to reduce the TC-Higgs mass from ${\cal O}$(1 TeV) to ${\cal O}$ (100 GeV). Clearly, as the four-fermion couplings reach the critical value for condensation, the expansion (\ref{Eq:Estimate}) becomes less and less accurate. On the other hand,  (\ref{Eq:Estimate}) shows unambiguously that four-fermion couplings much below the critical value cannot lower the scalar mass from values around $1$ TeV to values of ${\cal O}(100)$ GeV. Therefore, we conclude that only strongly-coupled ETC interactions can lower the mass of the TC-Higgs to the observed value of 125 GeV. This argument does not apply to TC theories with near-conformal dynamics, which are expected to feature a light scalar resonance even in isolation, {\em {\em i.e.}} with the ETC interactions switched off. 

As this discussion implies, there will be some amount of fine tuning involved in this scenario. However, the same applies to practically any BSM scenario which explains the relatively small mass of the observed Higgs boson. For example, in walking TC the fine tuning is associated with engineering the matter content or couplings so that the theory is only slightly within the critical region for the onset of spontaneous chiral symmetry breaking. In our case the amount of fine tuning can be quantified in a simple way as
\be
{\textrm{FT}}=\frac{M_H^2}{M_{H0}^2}.
\ee
The concrete values are on the level of few percents. For example, for $M_{H0}\simeq 1$ TeV, ${\rm FT}\simeq 1.6 \%$, whereas for $M_{H0}\sim 600$ GeV, ${\rm FT}\simeq 4.3\%$.

We end this section by evaluating the wave function renormalization $Z_H\equiv\Sigma_{HH}^\prime$. In order to compute the latter, we need
\begin{eqnarray}
\frac{d I_{XX}^{SS}}{d q^2} = 2\, J_{XX}+8\left(\frac{q^2}{4\, M_X^2}-1\right)J_{XX}^\prime\ .
\end{eqnarray}
We obtain
\begin{eqnarray}
Z_H &=& \frac{N\, y^2}{\left(1-N\, N_c\, G_{QqUt}^2\, {\cal I}_{UU}^{SS}\, {\cal I}_{tt}^{SS}\right)^2 \left(1-N\, G_{QQUU}\, I_{UU}^{SS}\right)^2}
\left[\frac{dI_{UU}^{SS}}{dq^2}
+\frac{N\, N_c\, G_{QqUt}^2\, \left({\cal I}_{tt}^{SS}\right)^2\, \left(I_{UU}^{SS}\right)^2}{\left(I_{tt}^{SS}\right)^2}\, \frac{dI_{tt}^{SS}}{dq^2}\right] \nonumber \\
&+&\frac{N\, y^2}{\left(1-N\, N_c\, G_{QqDb}^2\, {\cal I}_{DD}^{SS}\, {\cal I}_{bb}^{SS}\right)^2\left(1-N\, G_{QQDD}\, I_{DD}^{SS}\right)^2}
\left[\frac{dI_{DD}^{SS}}{dq^2}
+\frac{N\, N_c\, G_{QqDb}^2\, \left({\cal I}_{bb}^{SS}\right)^2 \left(I_{DD}^{SS}\right)^2}{\left(I_{bb}^{SS}\right)^2}\,\frac{dI_{bb}^{SS}}{dq^2}\right] \ .
\end{eqnarray}
Neglecting terms of order $M_X^2/\Lambda^2$, and using the fermion-mass equations, gives
\begin{eqnarray}
Z_H \simeq \frac{y^2}{M_Q^2}\left[
N\, M_U^2\, \frac{d I_{UU}^{SS}}{dq^2} + N\, M_D^2\, \frac{d I_{DD}^{SS}}{dq^2} + N_c\, M_t^2\, \frac{d I_{tt}^{SS}}{dq^2} + N_c\, M_b^2\, \frac{d I_{bb}^{SS}}{dq^2} \right] \ .
\end{eqnarray}
To a good approximation we may also ignore $J_{XX}^\prime/J_{XX}$, and set $J_{XX}\simeq 2K_{XY}\simeq 2K_{YX}$, where $X$ and $Y$ are $U$ and $D$ or $t$ and $b$. Using (\ref{Eq:v}), this gives
\begin{equation}
Z_H\simeq \frac{y^2}{\sqrt2\, G_F\, M_Q^2}\ .
\label{Eq:ZHapprox}
\end{equation}
\section{Couplings of the TC Higgs}
\label{Sec:couplings}
In this section we compute the coupling of the TC Higgs with the constituent techniquarks, the SM quarks, and the weak bosons. The computation of the TC-Higgs coupling to two photons will not be considered in this note.
\subsection{Coupling to fermions}
\begin{figure}[htb]
\includegraphics[width=4.5in]{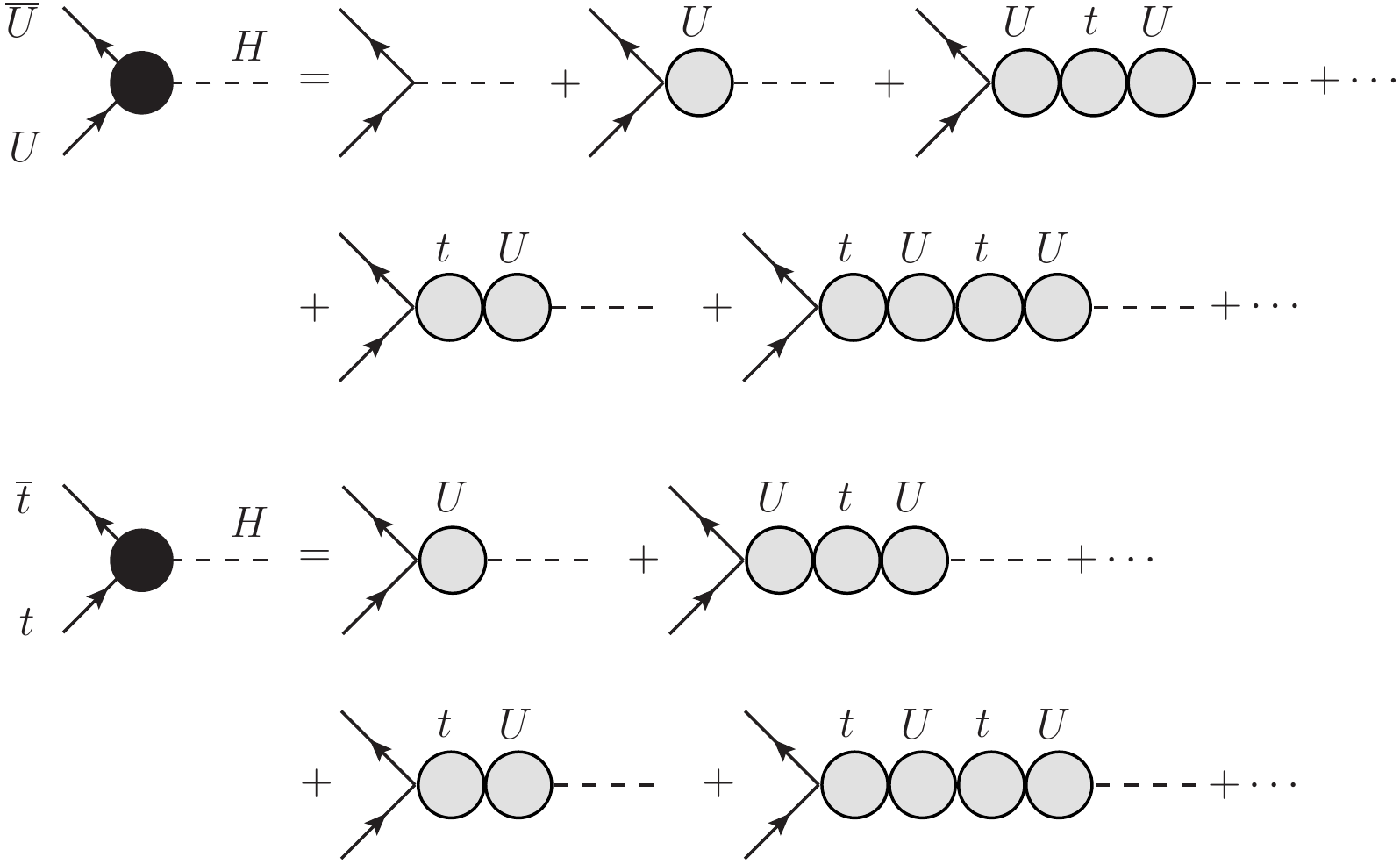}
\caption{Diagrams contributing to the $U\overline{U}H$ and $t\overline{t}H$ Yukawa vertices. The $D\overline{D}H$ and $b\overline{b}H$ vertices are just obtained by replacing $U$ and $t$ with $D$ and $b$ everywhere.}
\label{Fig:Hff}
\end{figure}
To leading order in the large-$N$ expansion, the diagrams contributing to the coupling of the TC-Higgs to fermions are shown in Fig.~\ref{Fig:Hff}. These lead to the effective Yukawa vertices
\begin{eqnarray}
{\cal L}_{\rm Yukawa} = - y_U\,  \overline{U}\, U\, H - y_D\,  \overline{D}\, D\, H - y_t\,  \overline{t}\, t\, H - y_b\,  \overline{b}\, b\, H \ ,
\end{eqnarray}
where
\begin{eqnarray}
y_U = \frac{1+N\, G_{QQUU}\, {\cal I}_{UU}^{SS}}{1-N\, N_c\, G_{QqUt}^2\, {\cal I}_{UU}^{SS}\, {\cal I}_{tt}^{SS}}\frac{y}{\sqrt{Z_H}}\ , \quad
y_t = \frac{N\, G_{QqUt}\, {\cal I}_{UU}^{SS}+N\, N_c\, G_{qqtt}\, G_{QqUt}\, {\cal I}_{tt}^{SS}\, {\cal I}_{UU}^{SS}}{1-N\, N_c\, G_{QqUt}^2\, {\cal I}_{UU}^{SS}\, {\cal I}_{tt}^{SS}}\frac{y}{\sqrt{Z_H}}\ ,  
\end{eqnarray}
and
\begin{eqnarray}
y_D = \frac{1+N\, G_{QQDD}\, {\cal I}_{DD}^{SS}}{1-N\, N_c\, G_{QqDb}^2\, {\cal I}_{DD}^{SS}\, {\cal I}_{bb}^{SS}}\frac{y}{\sqrt{Z_H}}\ , \quad
y_b = \frac{N\, G_{QqDb}\, {\cal I}_{DD}^{SS}+N\, N_c\, G_{qqbb}\, G_{QqDb}\, {\cal I}_{bb}^{SS}\, {\cal I}_{DD}^{SS}}{1-N\, N_c\, G_{QqDb}^2\, {\cal I}_{DD}^{SS}\, {\cal I}_{bb}^{SS}}\frac{y}{\sqrt{Z_H}}\ .
\end{eqnarray}
Neglecting terms of order $M_X^2/\Lambda^2$, and using equations (\ref{Eq:Ut}), (\ref{Eq:Db}), as well as the approximation (\ref{Eq:ZHapprox}), leads to
\begin{eqnarray}
y_U \simeq \left(\sqrt2\, G_F\right)^{1/2}\, M_U \ , \quad
y_t \simeq \left(\sqrt2\, G_F\right)^{1/2}\, M_t\ , \quad
\label{Eq:yUyt}
\end{eqnarray}
and
\begin{eqnarray}
y_D \simeq \left(\sqrt2\, G_F\right)^{1/2}\, M_D\ , \quad
y_b \simeq \left(\sqrt2\, G_F\right)^{1/2}\, M_b\ ,
\label{Eq:yDyb}
\end{eqnarray}
Therefore, the top and bottom Yukawa couplings are close to their SM values. We shall evaluate the Yukawa couplings numerically in Sec. \ref{Sec:Results}.
\subsection{Coupling to weak bosons}
\begin{figure}
\includegraphics[width=6in]{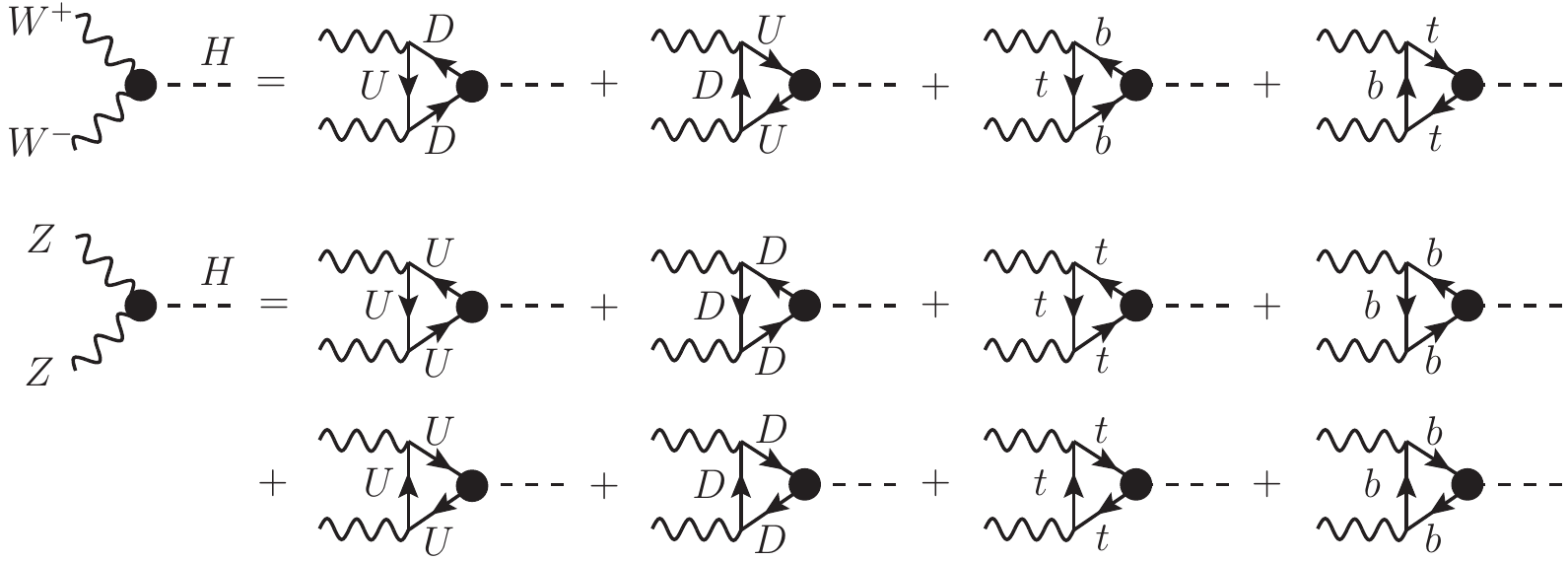}
\caption{Diagrams contributing to the coupling of the TC-Higgs with the $W$ boson (top) and the $Z$ boson (bottom). The Yukawa vertices are denoted by a black disk, and correspond to the diagrams of Fig. \ref{Fig:Hff}}
\label{Fig:HWW}
\end{figure}
To leading order in the large-$N$ expansion, the diagrams contributing to the coupling of the TC-Higgs to the $W$ and $Z$ boson are shown in Fig.~\ref{Fig:HWW}. These are given in terms of the effective Yukawa vertices computed above. At zero momentum, these diagrams lead to the effective vertices
\begin{equation}
{\cal L}_{HWW} = 2\, M_W^2\, \left(\sqrt2 G_F\right)^{1/2}\, a_W\, H\, W_\mu^+\, W^{-\mu}+M_Z^2\, \left(\sqrt2 G_F\right)^{1/2}\, a_Z\, H\, Z_\mu\, Z^\mu
\end{equation}
where
\begin{eqnarray}
a_W &=& 4\left(\sqrt2\, G_F\right)^{1/2}\left(N\, y_U\, M_U\, K_{UD}+N\, y_D\, M_D\, K_{DU}+N_c\, y_t\, M_t\, K_{tb}+N_c\, y_b\, M_b\, K_{bt}\right)\ , \nonumber \\
a_Z &=& 2\left(\sqrt2\, G_F\right)^{1/2}\left(N\, y_U\, M_U\, J_{UU}+N\, y_D\, M_D\, J_{DD}+N_c\, y_t\, M_t\, J_{tt}+N_c\, y_b\, M_b\, J_{bb}\right)\ .
\end{eqnarray}
We may use the approximation $J_{XY}\simeq 2K_{XY}$, which is only valid as long as the isospin mass splitting is small, and thus for phenomenologically viable $M_U$ and $M_D$. However, we can use this approximation also for $M_t$ and $M_b$, as the corresponding contributions to $a_W$ and $a_Z$ are much smaller.
Then, using (\ref{Eq:yUyt}) and (\ref{Eq:yDyb}), as well as (\ref{Eq:v}), gives
\begin{eqnarray}
a_Z \simeq a_W \simeq  \sqrt2\, G_F\times 4 \left(N\, M_U^2\, K_{UD}+N\, M_D^2\, K_{DU}+N_c\, M_t^2\, K_{tb}+N_c\, M_b^2\, K_{bt}\right)\simeq 1\ .
\end{eqnarray}
The $g_{HWW}$ and $g_{HZZ}$ couplings are therefore close to their SM values. We shall evaluate numerically $a_W$ and $a_Z$ in Sec. \ref{Sec:Results}.
\section{Electroweak parameters}
\label{Sec:ewparams}
\begin{figure}
\includegraphics[width=4.5in]{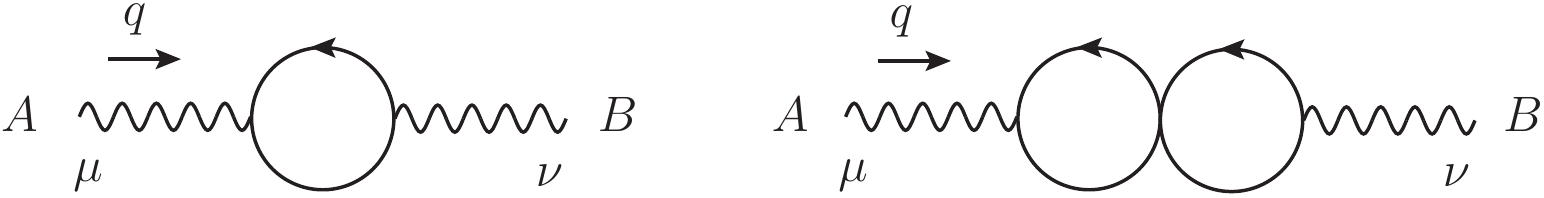}
\caption{One-loop and two-loop diagrams giving the dominant contributions to the $g^{\mu\nu}$ part of the electroweak VPAs. See text for details.}
\label{Fig:ST}
\end{figure}

The general form of the electroweak vacuum polarisation amplitude (VPA) is
\begin{eqnarray}
\Pi_{AB}^{\mu\nu}(q)=\Pi_{AB}(q^2)\left(g^{\mu\nu}-\frac{q^\mu q^\nu}{q^2}\right)\ .
\end{eqnarray}
As explained in Sec. \ref{Sec:Masses}, in order to recover a fully transverse result, all contributions have to be taken into account, including tree-level exchanges of Goldstone bosons. However $\Pi_{AB}(q^2)$ can be more easily extracted from the $g^{\mu\nu}$ part, which requires computing less diagrams. For instance, the four-fermion operators not included in $\Delta{\cal L}_{\rm ETC}$ only contribute indirectly to $g^{\mu\nu}$, by affecting the fermion masses. Therefore, their contribution to $\Pi_{AB}(q^2)$ can be extracted from one-loop diagrams. On the other hand, the operators contained in $\Delta{\cal L}_{\rm ETC}$ contribute directly to the $g^{\mu\nu}$ part of the VPAs, and a chain of fermion loops must be computed to obtain the full leading-$N$ contribution. However the fermion bubbles with external vectors are only logarithmically divergent, and each loop brings a suppression factor of the order of $M_X^2/{\cal M}^2$. Therefore, the full contribution to $\Pi_{AB}(q^2)$ from the operators contained in $\Delta{\cal L}_{\rm ETC}$ is of the form $\Pi_{AB}(q^2)={\rm two\ loop}\times\left[1+{\cal O}(M_X^2/{\cal M}^2)\right]\simeq {\rm two\ loop}$. As a consequence, in order to evaluate $\Pi_{AB}(q^2)$ we only need to compute one-loop and two-loop diagrams, as shown in Fig. \ref{Fig:ST}: the one-loop diagrams give the contribution from the four-fermion operators not contained in $\Delta{\cal L}_{\rm ETC}$ (through modified fermion masses), whereas the two-loop diagrams give, to an excellent approximation, the contribution from the operators contained in $\Delta{\cal L}_{\rm ETC}$.

The $S$ and $T$ parameters are defined by
\begin{eqnarray}
&& S=\frac{16\pi}{g\, g^\prime}\, \Pi_{W^3 B}^\prime(0) \ , \\
&& T = \frac{1}{\alpha\, M_W^2}\left[\Pi_{W^3 W^3}(0)-\Pi_{W^+ W^-}(0)\right] \ ,
\end{eqnarray}
where $\alpha$ is the electromagnetic coupling evaluated at the $Z$ pole. We define standard units for $S$ and $T$:
\begin{equation}
S_0\equiv \frac{1}{6\pi}\simeq 0.05\ , \quad
T_0 \equiv \frac{\sqrt2\, G_F\, M_t^2}{16\pi^2\, \alpha}\simeq 0.40\ .
\label{Eq:S0T0}
\end{equation}
Now we shall consider the different contributions to $S$ and $T$ from the four-fermion operators. Unlike done so far, we will neglect the bottom mass, except, of course, in $\log{\cal M}^2/M_b^2$. The building-block integrals for evaluating the VPAs are
\begin{eqnarray}
&& I_{XY}^{RR}\, g^{\mu\nu} + q^\mu q^\nu\ {\rm terms}  = I_{XY}^{LL}\, g^{\mu\nu} + q^\mu q^\nu\ {\rm terms} \nonumber \\
&& \equiv i \int \frac{d^4 k}{(2\pi)^4}\, {\rm Tr}\, \gamma^\mu\, P_L\, \frac{i\left(\slashed{k}+M_Y\right)}{k^2-M_Y^2+i\, \varepsilon} \gamma^\nu\, P_L
\frac{i\left(\slashed{k}+\slashed{q}+M_X\right)}{(k+q)^2-M_X^2+i\, \varepsilon}
\end{eqnarray}
and
\begin{eqnarray}
&& I_{XY}^{RL}\, g^{\mu\nu} + q^\mu q^\nu\ {\rm terms}  = I_{XY}^{LR}\, g^{\mu\nu} + q^\mu q^\nu\ {\rm terms} \nonumber \\
&& \equiv i \int \frac{d^4 k}{(2\pi)^4}\, {\rm Tr}\, \gamma^\mu\, P_L\, \frac{i\left(\slashed{k}+M_Y\right)}{k^2-M_Y^2+i\, \varepsilon} \gamma^\nu\, P_R
\frac{i\left(\slashed{k}+\slashed{q}+M_X\right)}{(k+q)^2-M_X^2+i\, \varepsilon} \ .
\end{eqnarray}
In order to compute $S$ and $T$ we need the integrals
\begin{align}
 \left(I_{XY}^{RR}\right)_{q^2=0} = \left(I_{XY}^{LL}\right)_{q^2=0} = -2\left(K_{XY}\, M_X^2+K_{YX}\, M_Y^2\right) \ , \quad
\left(I_{XY}^{RL}\right)_{q^2=0} = \left(I_{XY}^{LR}\right)_{q^2=0} = 2\, J_{XY}\, M_X\, M_Y \ ,  \nonumber \\
\end{align}
and the derivatives
\begin{align}
 \left(\frac{d I_{XX}^{RR}}{d q^2}\right)_{q^2=0} = \left(\frac{d I_{XX}^{LL}}{d q^2}\right)_{q^2=0} = 4\, L_{XX}-2\, J_{XX}^\prime\ , \quad
\left(\frac{d I_{XX}^{RL}}{d q^2}\right)_{q^2=0} = \left(\frac{d I_{XX}^{LR}}{d q^2}\right)_{q^2=0} = 2\, J_{XX}^\prime \ ,
\end{align}
together with (\ref{Eq:RelInt}).
%
%
The one-loop contributions are simple and reproduce the usual results, namely
\begin{eqnarray}
S_{\rm one\ loop} &=& 16\pi\left[
\frac{N}{2}\left(J_{UU}^\prime+J_{DD}^\prime\right)+\frac{N_c}{2}\left(J_{tt}^\prime+J_{bb}^\prime\right)+\frac{N_c}{18}\left(J_{tt}-J_{bb}\right)\right]-S_{\rm SM} \nonumber \\
&=&N\, S_0\left[1+{\cal O}\left(\frac{M_Q^2}{\Lambda^2}\right)+{\cal O}\left(\frac{M_t^2}{{\cal M}^2}\right)\right] \ ,
\end{eqnarray}
for the $S$-parameter and
\begin{eqnarray}
T_{\rm one\ loop} &=& \frac{4\sqrt2\, G_F}{\alpha}\left[
N\left(K_{UD}-\frac{J_{UU}}{2}\right)M_U^2+N\left(K_{DU}-\frac{J_{DD}}{2}\right)M_D^2+N_c\left(K_{tb}-\frac{J_{tt}}{2}\right)M_t^2\right] -T_{\rm SM} \nonumber \\
&=& N\, T_0\, \frac{M_U^4-M_D^4-2\, M_U^2\, M_D^2\, \log\left(M_U^2/M_D^2\right)}{\left(M_U^2-M_D^2\right)M_t^2}\left[1+{\cal O}\left(\frac{M_Q^2}{\Lambda^2}\right)+{\cal O}\left(\frac{M_t^2}{{\cal M}^2}\right)\right]
\end{eqnarray}
for the $T$ parameter.
%
%
Then we consider the leading contributions, in the large-$N$ expansion, from the operators contained in $\Delta{\cal L}_{\rm ETC}$. As argued above, these are dominantly given by two-loop diagrams. We label them according to the corresponding product of ETC couplings. The formulas are collected in Appendix \ref{app:twoloopST}.

\section{Numerical results}\label{Sec:Results}
The model parameters are $g$, $g^\prime$, $y$, $M_Q$, $M$, $\Lambda$, ${\cal M}$, $g_{QQ}$, $g_{qq}$, $g_{Qq}$, $g_{UU}$, $g_{DD}$, $g_{UD}$, $g_{tt}$, $g_{bb}$, $g_{tb}$, $g_{Ut}$, $g_{Dt}$, $g_{Ub}$, $g_{Db}$, and $N$. The Yukawa coupling $y$ disappears after renormalization, whereas $M$ can be exchanged for the dynamical mass $M_{H0}$. We trade $g$, $g^\prime$, $M_Q$, $M_{H0}$, $g_{Ut}$, and $g_{Db}$ for the experimental values of  $\alpha$, $G_F$, $M_Z$, $M_H$, $M_t$, and $M_b$, respectively. Therefore, we end up with the free variables $\Lambda$, ${\cal M}$, $g_{QQ}$, $g_{qq}$, $g_{Qq}$, $g_{UU}$, $g_{DD}$, $g_{UD}$, $g_{tt}$, $g_{bb}$, $g_{tb}$, $g_{Dt}$, $g_{Ub}$, and $N$. We may obtain an estimate of $\Lambda$ by scaling up the corresponding quantity from QCD. Putting together the Pagels-Stokar equation and the NJL formula for the mass of the $\sigma$ meson, leads to the equation
\begin{eqnarray}
f_\pi^2 = \frac{N_c}{16\pi^2}\, m_\sigma^2\, {\overline \log}\, \frac{\Lambda_{\rm QCD}^2}{m_\sigma^2/4}\ ,
\end{eqnarray}
where $\Lambda_{QCD}$ here is the mass scale of the non-Goldstone states and should not be confused with the standard one defined in the literature (of the order of 200 MeV). We may solve for $\Lambda_{\rm QCD}$, and use the scaling law
\begin{equation}
\Lambda = \sqrt{\frac{N_c}{N}}\, \frac{F_\Pi}{f_\pi}\, \Lambda_{\rm QCD}\ ,
\end{equation}
where $F_\Pi=(\sqrt2\, G_F)^{-1/2}\simeq 246$ GeV. Setting $f_\pi\simeq 93$ MeV, $m_\sigma\simeq 441$ MeV \cite{Caprini:2005zr}, and $N_c=3$, gives
\begin{equation}
\Lambda\simeq\left\{
\begin{array}{ll}
2.7\ {\rm TeV} & N=4 \\
2.2\ {\rm TeV} & N=6\ .
\end{array}
\right.
\end{equation}
We take these as an input to the numerical calculation. The ETC scale ${\cal M}$ remains as a free parameter, and we consider values increasing from ${\cal M}=\Lambda$. Then, we perform a random scan over the couplings $g_{QQ}$, $g_{qq}$, $g_{Qq}$, $g_{UU}$, $g_{DD}$, $g_{UD}$, $g_{tt}$, $g_{bb}$, $g_{tb}$, $g_{Dt}$, $g_{Ub}$. 

The conditions of sub-criticality (\ref{Eq:Constr1}) and (\ref{Eq:Constr2}) imply direct bounds on the ETC couplings. Using (\ref{Eq:MassOp}), we obtain 
\begin{eqnarray}
g_{QQ} g_{UU} < \frac{{\cal M}^2}{\Lambda^2}4\pi^2\ , \quad
g_{QQ} g_{DD} < \frac{{\cal M}^2}{\Lambda^2}4\pi^2\ , \quad
g_{qq} g_{tt} < 4\pi^2\ , \quad
g_{qq} g_{bb} < 4\pi^2 \ ,
\end{eqnarray}
and
\begin{eqnarray}
g_{Qq} g_{Ut} < \frac{{\cal M}}{\Lambda}\,   \frac{4\pi^2}{\sqrt{N\, N_c}}\ , \quad
g_{Qq} g_{Db} < \frac{{\cal M}}{\Lambda}\,  \frac{4\pi^2}{\sqrt{N\, N_c}} \ .
\end{eqnarray}
To further restrict the values of the parameters we consider the following:
\begin{itemize}
\item Taking $g_{UU}\neq g_{DD}$ causes large corrections to $T$, as one can see by adding together (\ref{Eq:TUUUU}), (\ref{Eq:TDDDD}) and (\ref{Eq:TUUDD}). To avoid this we impose $g_{UU}=g_{DD}$. One can think of this relation as arising from an approximate custodial symmetry in the ETC Lagrangian. With this motivation we also set $g_{tt}=g_{bb}$.
\item The coupling $g_{UD}$ gives an unbalanced large and negative contribution to $T$, and thus we choose to set $g_{UD}=0$. As for $g_{UU}=g_{DD}$ and $g_{tt}=g_{bb}$, also this relation can be thought of arising from a custodial symmetry in the ETC sector. This prompts us to set $g_{tb}=0$, although the corresponding contribution to the oblique parameters is negligible. 
\item The mass $M_U$ is always larger than $M_D$ because of the contribution from $G_{Ut}$ and the imposed relation $g_{UU}=g_{DD}$. From (\ref{Eq:TUtUt}) we see that $|g_{Ut}|$ cannot be too large, or else a large and positive contribution to $T$ appears. Since $G_{Ut}\equiv g_{Qq}\, g_{Ut}/{\cal M}^2$ is fixed by the top mass, this means that $|g_{qQ}|$ cannot be too small. We impose $\pi/3 \leq |g_{Qq}| \leq 2\pi$.
\item Equation (\ref{Eq:TDtDt}) shows that a positive contribution to $T$ can be balanced by increasing $|g_{Dt}|$ which, by a similar argument as above for $|g_{qQ}|$, should not be too small. Thus we require $\pi/3 \leq |g_{Dt}| \leq 2\pi$. This implies also a negative contribution to $S$, as shown by (\ref{Eq:SDtDt}).
\item The coupling $|g_{Ub}|$ should be large enough to give a negative contribution to $S$, see Eq. (\ref{Eq:SUb}). Also note that this coupling does not contribute to $T$, see Eq. (\ref{Eq:TUb}). We require $\pi/3 \leq |g_{Ub}| \leq 2\pi$.
\item As we have discussed earlier, large ETC couplings are required to lower the Higgs mass to the observed value. This is particularly true for $g_{QQ}$ and $g_{UU}=g_{DD}$, and thus we require these to be large enough by imposing $\pi/3 \leq |g_{QQ}|, |g_{UU}|=|g_{DD}| \leq 2\pi$.
\end{itemize}

Therefore, we scan the parameter space by assigning random values for the couplings within the following ranges:
\begin{eqnarray}
\pi/3\le |g_{QQ}|, |g_{UU}|, |g_{Qq}|, |g_{Dt}|, |g_{Ub}| \le 2\pi, \,\, -2\pi\le |g_{qq}|, |g_{tt}|\le 2\pi,
\label{Eq:scanrange}
\end{eqnarray}
and we also set 
\begin{eqnarray}
g_{DD}=g_{UU}\ , \quad g_{bb} = g_{tt}\ , \quad g_{UD}=0\ , \quad g_{tb} = 0\ .
\label{Eq:Cust}
\end{eqnarray}
\begin{figure}
\includegraphics[width=.45\textwidth]{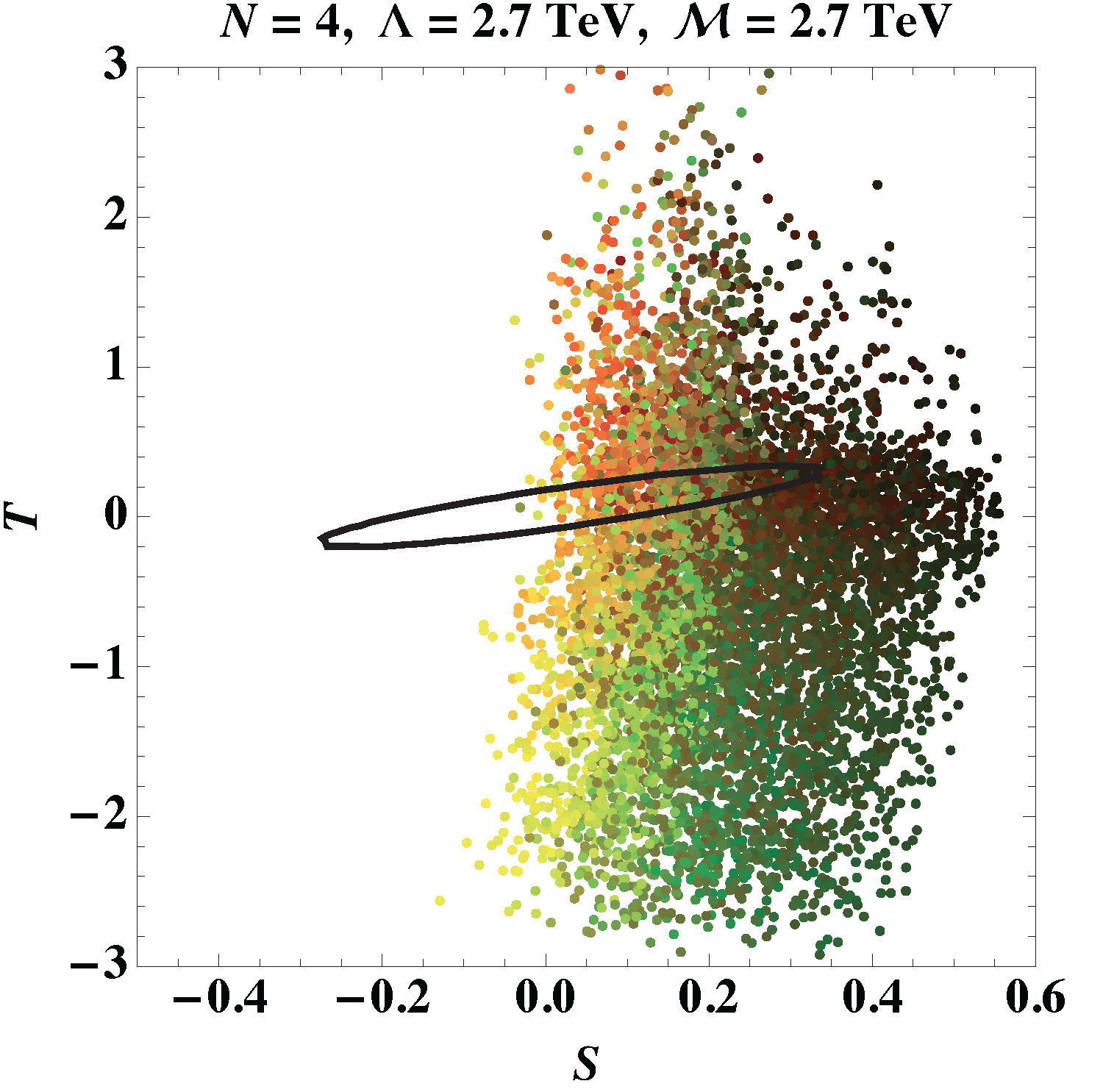}
\includegraphics[width=.45\textwidth]{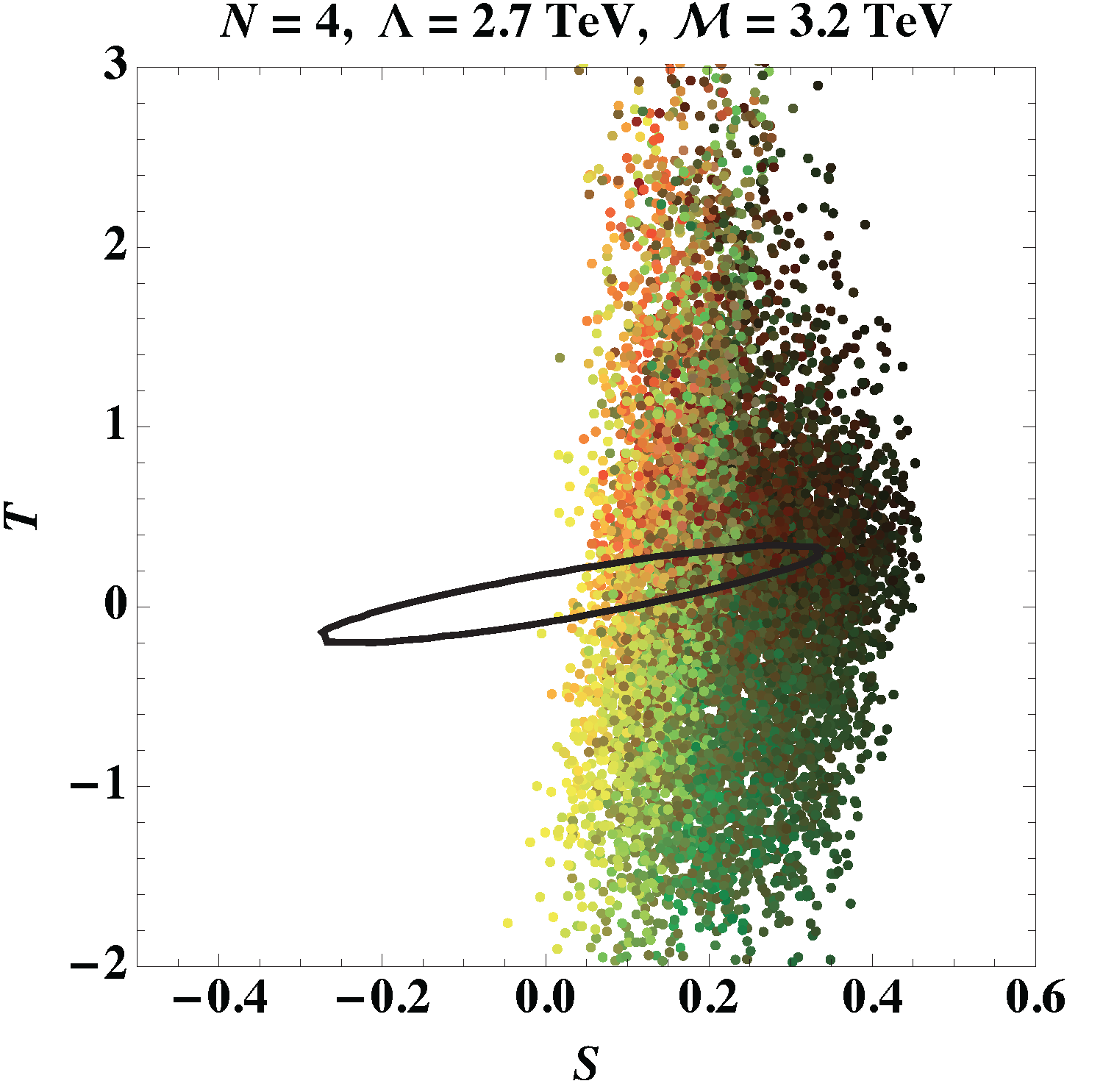}

\includegraphics[width=.45\textwidth]{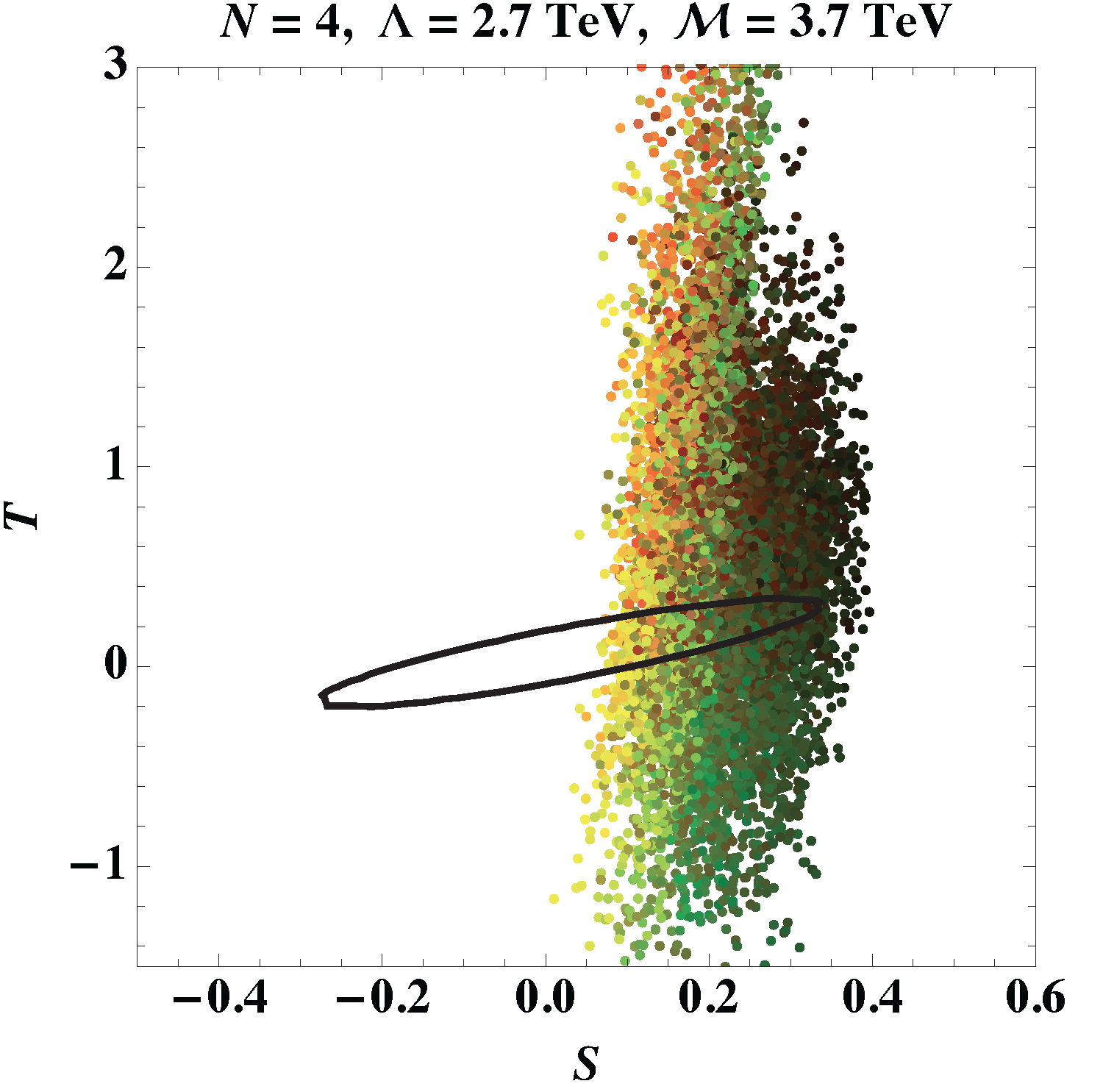}
\includegraphics[width=.45\textwidth]{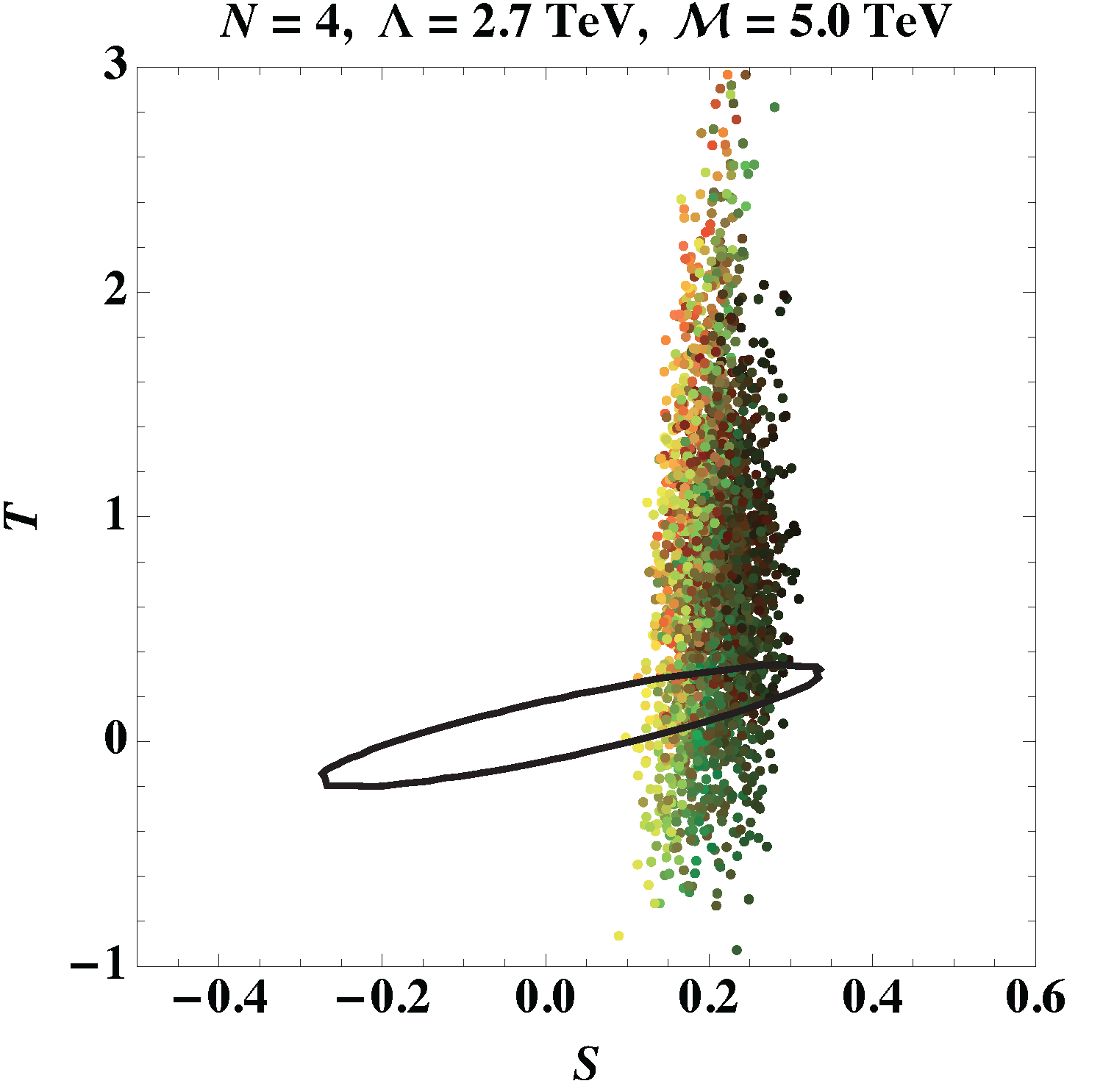}
\caption{Result over the random scan of the model parameter space for $N=4$ and $\Lambda=2.7$ TeV. The four figures correspond to $M=2.7$, $3.2$, $3.7$  and 5 TeV as indicated in the figure labels. The details of the scan are explained in the text and all points shown in the figure correspond to the Higgs mass 125 GeV. The shown points have the dynamical Higgs mass between fractions 0.5 to 1.25 of the scaled up mass of the sigma meson.}
\label{Fig:N4}
\end{figure}

\begin{figure}
\includegraphics[width=0.45\textwidth]{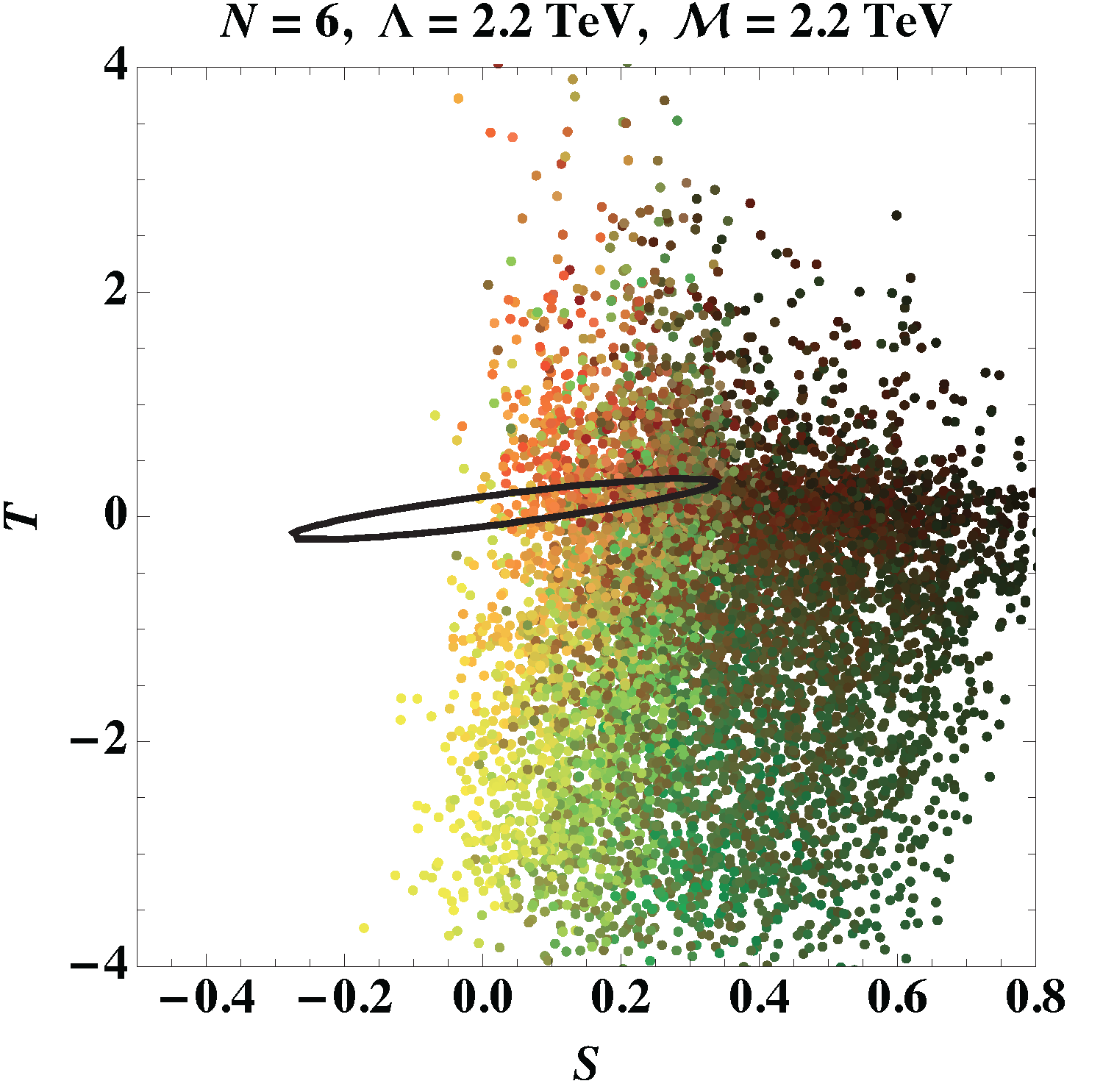}
\includegraphics[width=0.45\textwidth]{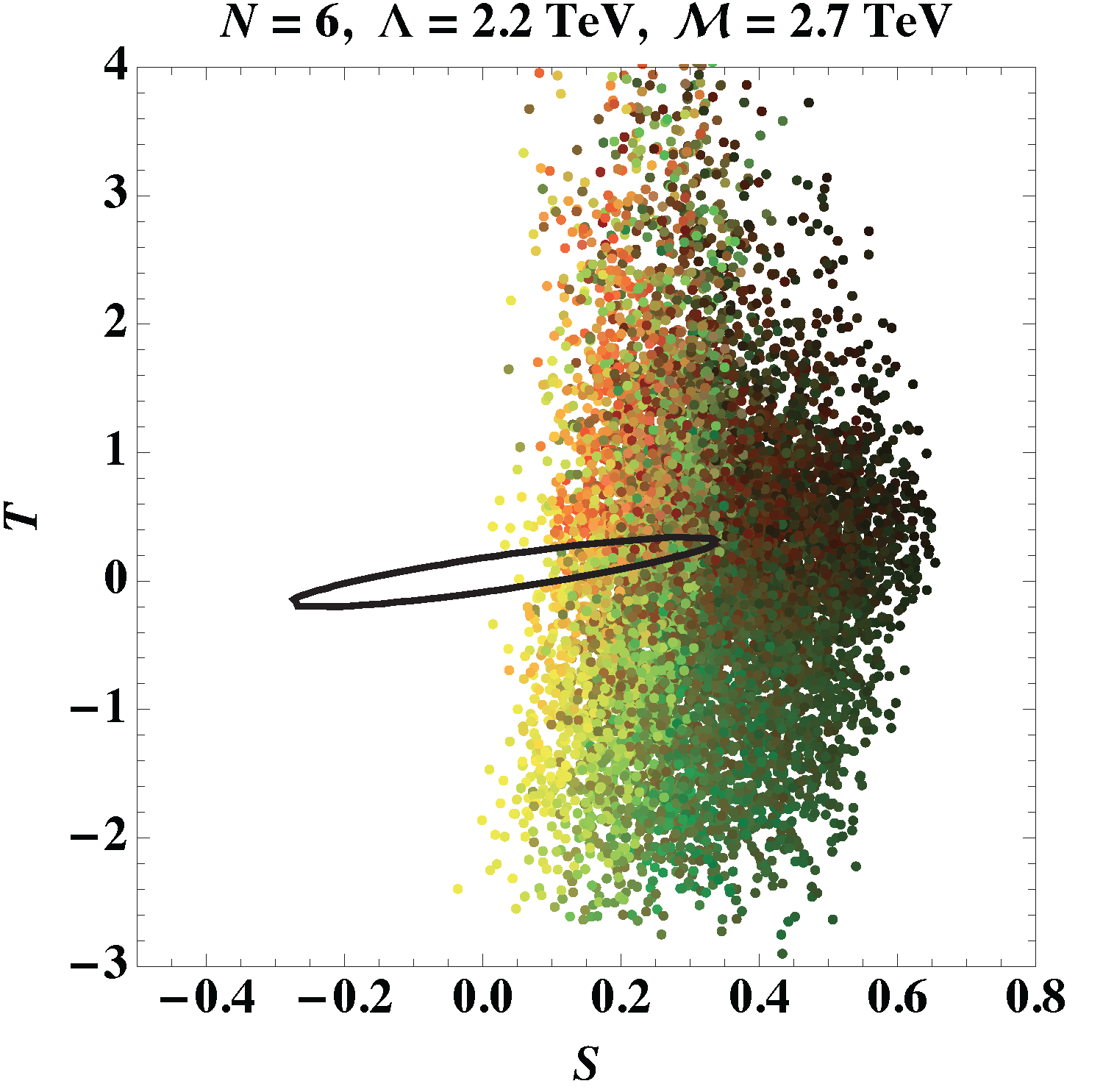}

\includegraphics[width=0.45\textwidth]{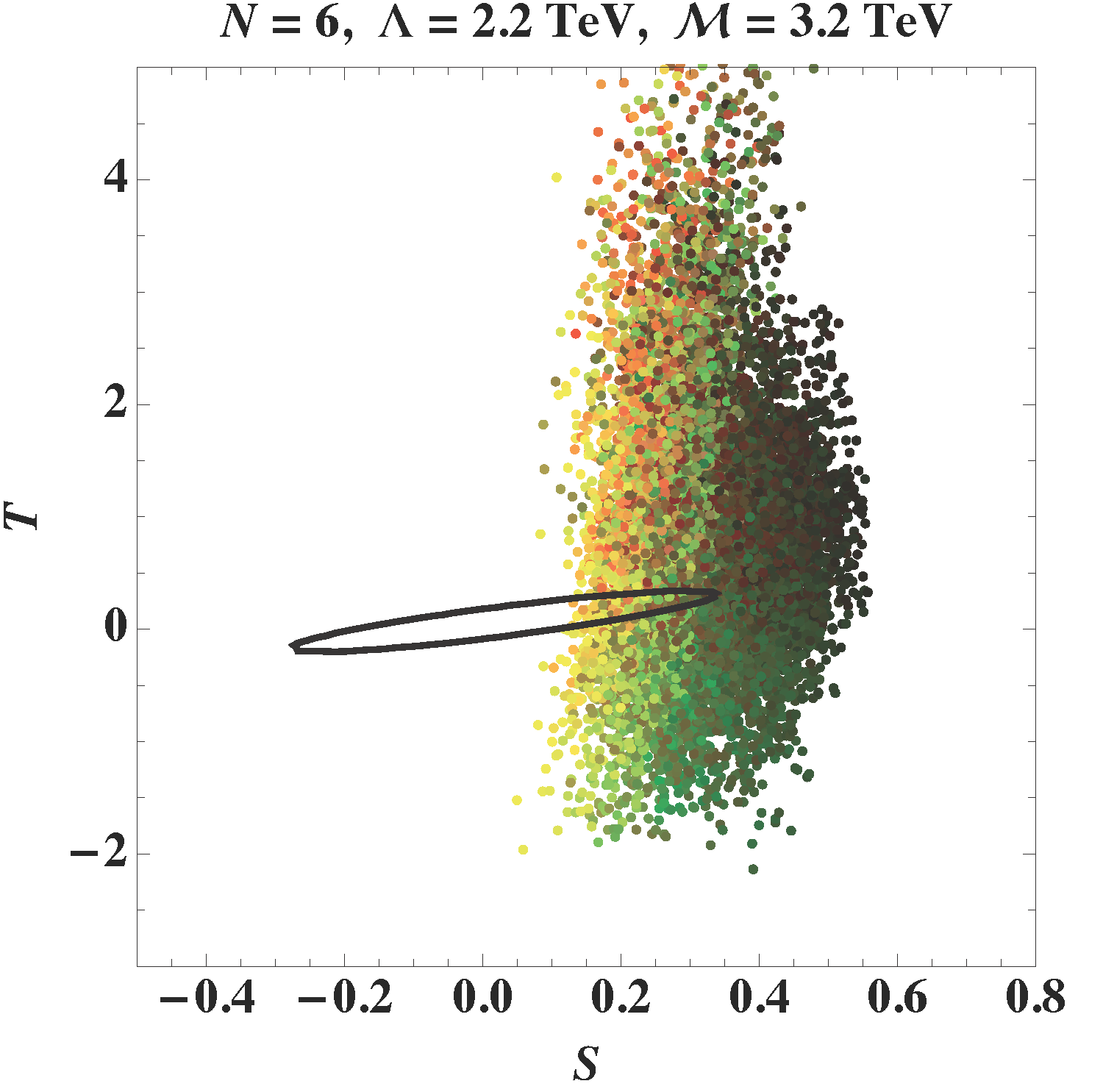}
\includegraphics[width=0.45\textwidth]{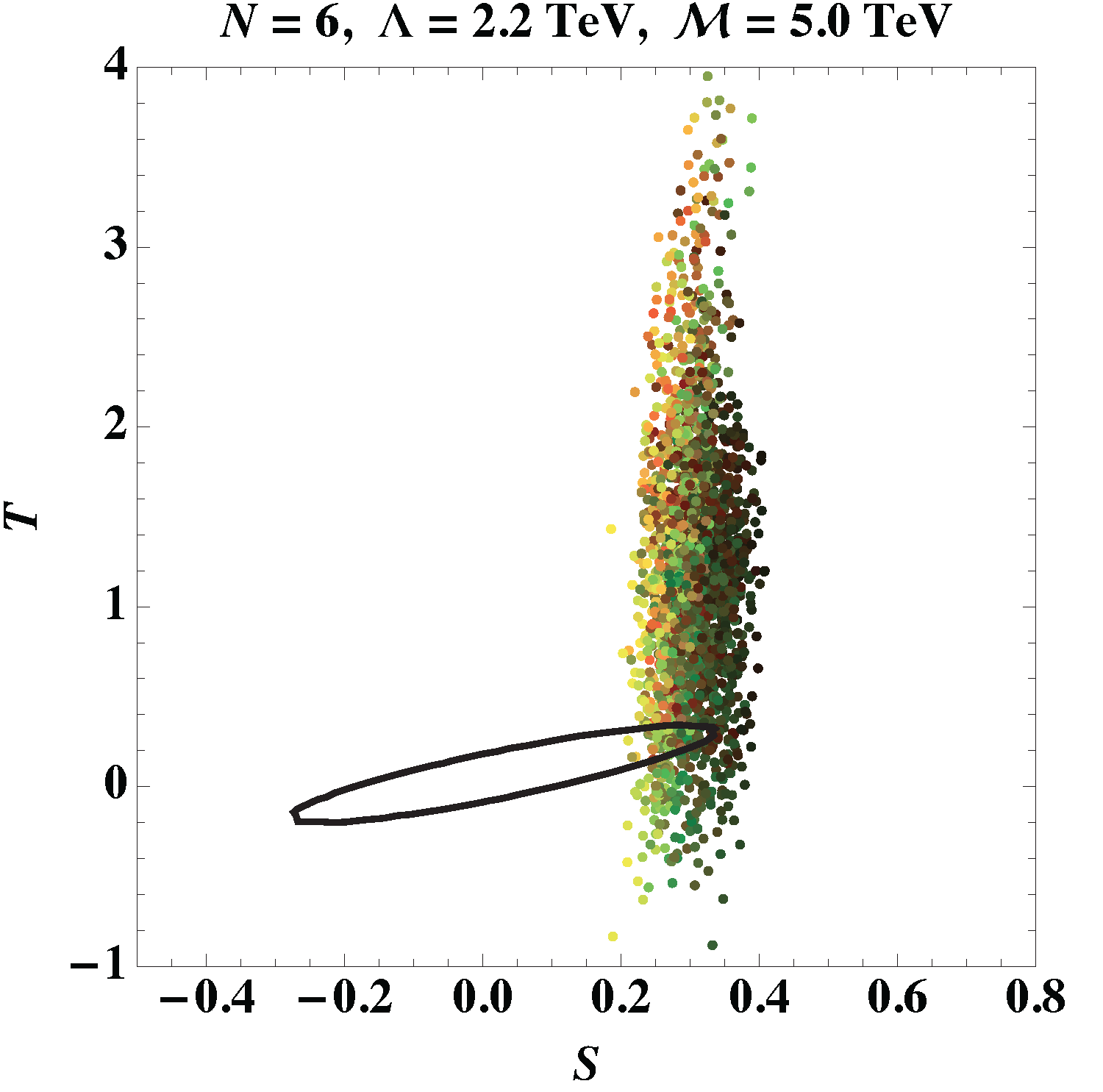}
\caption{Result over the random scan of the model parameter space for $N=6$ and $\Lambda=2.2$ TeV. The four figures correspond to $M=2.2$, $2,7$, $3.2$  and 5 TeV, respectively.}
\label{Fig:N6}
\end{figure}

For each generated data point we compute masses, the couplings of the Higgs particle, and evaluate the $S$ and $T$ parameters. As shown in section \ref{Sec:couplings}, the Higgs couplings are expected to be close to their SM values. Numerical computations show that this is true within $\sim 10\%$ for any value of the ETC couplings and mass. In particular, the Higgs coupling to the weak boson is always slightly larger than the SM value, the top Yukawa is slightly smaller, whereas the bottom Yukawa is standard within a few per cents.

For each considered value of $\Lambda$ and ${\cal M}$, we sampled 25 000 points satisfying (\ref{Eq:scanrange}) and (\ref{Eq:Cust}), and plotted these in the $S,T$ plane, together with the experimentally viable $3\sigma$ contour. The dynamical mass $M_{H0}$ is of course highly dependent on the ETC couplings. In Figs. \ref{Fig:N4}, for $N=4$, and \ref{Fig:N6}, for $N=6$, we only show those points for which $M_{H0}$ is between the fractions 0.5 and 1.25 of the scaled up mass of the sigma meson:
\begin{equation}
0.5\, m_\sigma\, \frac{F_\Pi}{f_\pi}\,\sqrt{\frac{N_c}{N}} < M_{H0} < 1.25\, m_\sigma\, \frac{F_\Pi}{f_\pi}\,\sqrt{\frac{N_c}{N}} \ .
\end{equation}
The reason to allow for smaller values of $M_{H0}$ is to account for possible effects from walking dynamics. The color coding of the points shown in the figure is related to the different couplings as follows: The bigger the value of $|g_{Dt}|$, the greener the point, and the larger the value of $|g_{Ub}|$, the redder the point. Furthermore, the larger the value of $|g_{Qq}|$ the darker the shade of the color, with a clear border at $|g_{Qq}|=3$. Larger values of $|g_{Ub}|$ imply smaller $S$ whereas larger values of $|g_{Dt}|$ imply smaller $S$ and also smaller $T$. Therefore, as we can observe from the figure, the redder points are towards the top-left of the plot while the greener points are towards the bottom.

\begin{figure}
\includegraphics[width=0.45\textwidth]{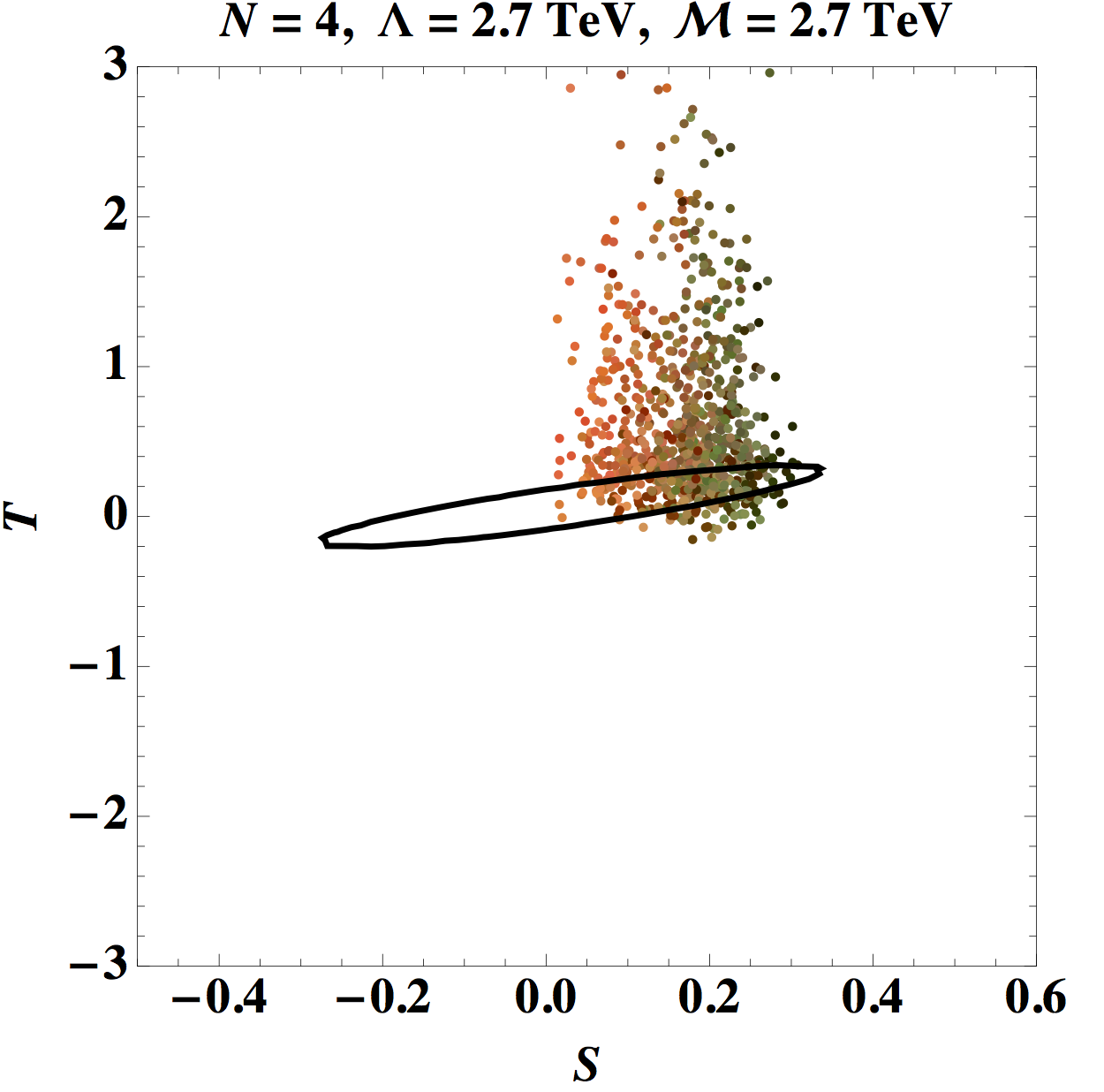}
\includegraphics[width=0.45\textwidth]{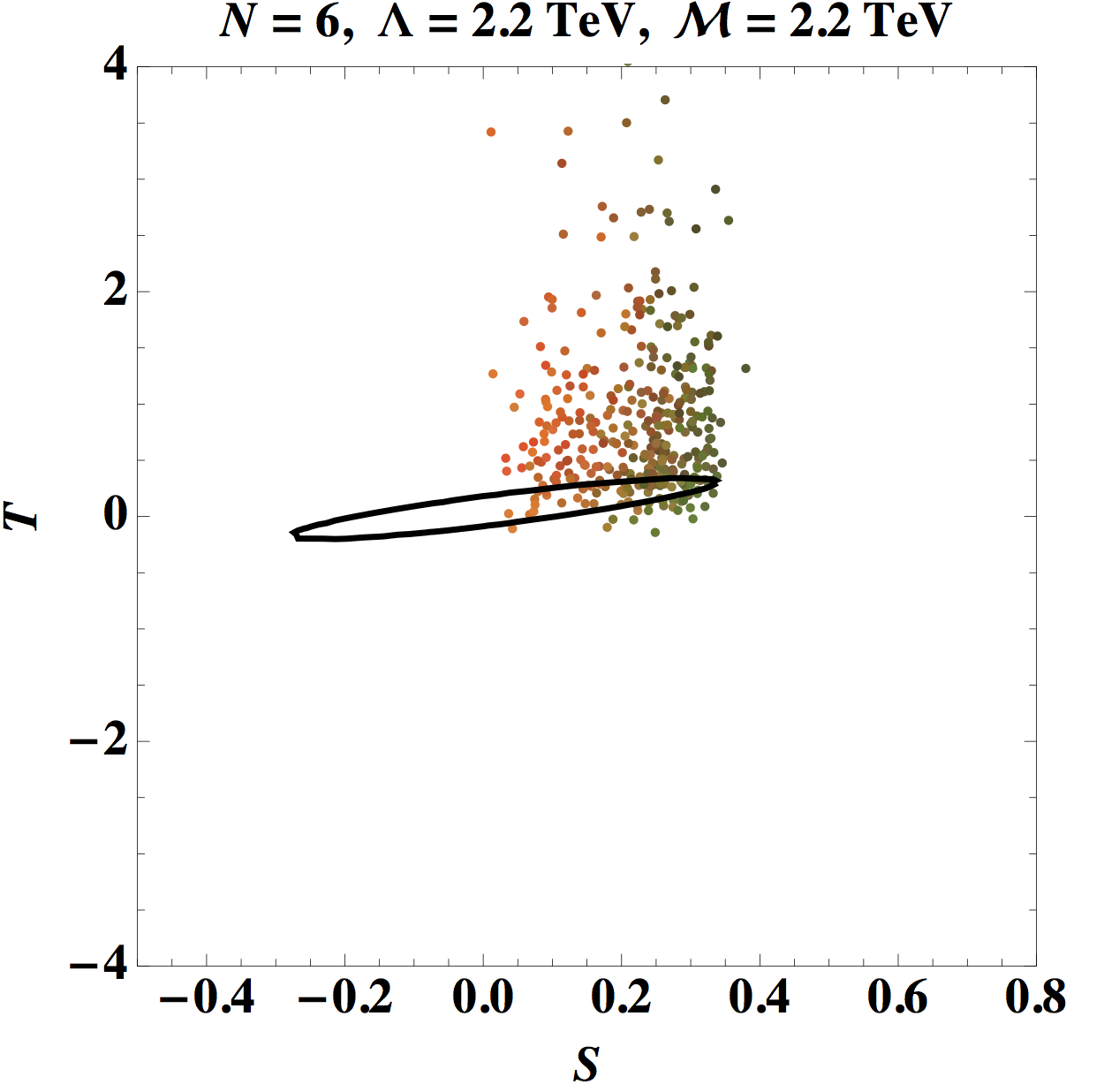}
\caption{The plots on the top-left corner in Figs. \ref{Fig:N4} and \ref{Fig:N6} are reproduced here with the additional constraints $|g_{Dt}| < 2.5$ and $|g_{Qq}| < 3.5$ ($|g_{Qq}| < 2.0$), for $N=4$ ($N=6$).}
\label{Fig:Restricted}
\end{figure}

For small values of ${\cal M}$ the green points are disfavoured. This is because $T$ is too large and negative. The favoured points on the other hand are orange, corresponding to approximately equal values of $|g_{Dt}|$ and $|g_{Ub}|$.  Increasing the ETC scale ${\cal M}$ makes the mass splitting $M_U-M_D$ larger. In the gap equations this splitting originates from $G_{QqUt}$, the latter being fixed by the top quark mass. The large $M_U-M_D$ splitting implies a large positive one-loop contribution to the $T$ parameter, and this needs to  be balanced by a negative contribution. Such a balancing contribution can only arise from $g_{Dt}$, as shown by (\ref{Eq:TDtDt}), and therefore for large ${\cal M}$ the favoured points are the greener ones.

Overall, for large values of ${\cal M}$, {\em e.g.}. 5 TeV, there are less points since larger values of $g_{QQ}g_{UU}$ and $g_{QQ}g_{DD}$ are required to have sufficient reduction for the Higgs mass. In other words, the portion of the sampled points shown in the figure shrinks as ${\cal M}$ increases. Finally, we note that varying $\Lambda$ within, say, $20\%$ range around the values we have used does not change the results qualitatively. For slightly smaller $\Lambda$ one needs slightly larger values of $g_{QQ}g_{UU}$ and $g_{QQ}g_{DD}$ to have correct reduction in the Higgs mass. However, the values of $S$ and $T$ are only little affected by these couplings in the limit $g_{UU}=g_{DD}$ which we have imposed.

We note that $T$ can take on values from a very large interval, making the agreement with experiment challenging. On the one one hand this is due to the fact that the natural size $T_0$, in (\ref{Eq:S0T0}), is relatively large, unlike $S_0$ which is small. On the other hand this occurs because we have allowed for a wide range of ETC couplings. To illustrate this, we plot in Fig. \ref{Fig:Restricted} points satisfying $|g_{Dt}| < 2.5$ and $|g_{Qq}| < 3.5$ ($|g_{Qq}| < 2.0$), for $N=4$ ($N=6$) and ${\cal M}=\Lambda$. We see that only a small portion of the original points are selected, and the electroweak parameters can take on values from much smaller intervals. In general the scatter plots in Figs. \ref{Fig:N4} and \ref{Fig:N6} can be considered as instructions on how to build an ETC theory satisfying the experimental constraints from precision electroweak observables. We leave for future work the more thorough analyses highlighting different domains and their relevance for microscopic ETC model building.
\section{Conclusions}
\label{Sec:checkout}
In this paper we have considered a realistic effective model where the electroweak scale and the masses of the weak gauge bosons arise dynamically from a TC force. We assume that the technifermions  are coupled with the SM matter fields via four-fermion interactions, low-energy remnants of an ETC theory, and classified all four-fermion operators based on the quantum numbers of the exchanged ETC gauge bosons.  We have demonstrated that the lightest scalar resonance can be as light as 125 GeV, and has SM-like couplings with the particles of the SM. We have also evaluated the contribution of the four fermion operators to the oblique electroweak parameters, and demonstrated that the model is viable in some region of the parameter space. This instructs us on how to build an ETC theory satisfying the experimental constraints on the precision electroweak parameters.

Observables have been computed in the large-$N$ scheme, where $N$ is the dimension of the  technifermion representation under the TC gauge group, and this has allowed us to obtain rigorous and robust results. One of our main conclusions is that in models of this type, in order to obtain a light composite Higgs, the ETC theory needs to be strongly coupled. We considered the case where ETC couplings were nevertheless required to be subcritical, and the success of this setup to produce a light scalar necessarily implies some amount of fine tuning. With the scales relevant for strong dynamics we considered, this fine tuning is on the level of few percents.

In addition to the phenomenological results, we have outlined features of the computation which are important for theories where multiple cutoff scales are introduced and gauge invariance needs to be maintained. These results are expected to have applications for a range of models involving strong dynamics. 

The results we have obtained here provide a solid benchmark model to investigate in light of current and future collider data. One can think of several further prospects which can be pursued: For example, additional composite states with masses of ${\cal O}$(1 Tev) are expected to appear in the spectrum, and their properties and couplings could be analysed in more detail. Also, the momentum dependence of the couplings of the scalar boson to the SM fields should be studied in more detail, as they could provide an important window into the possible composite nature of the Higgs boson.

Finally, our results can be applied to strengthen the analysis of the phenomenological implications of the lattice results on new strong dynamics. On the one hand, the lattice computations are performed on strong dynamics in isolation, and these first principle analyses provide valuable input on the scales appearing in the model setup we have analysed in this paper. On the other hand, our results for the corrections to the scalar mass from external four fermion interactions, when implemented on the lattice results of the scalar particle, can help in estimating the applicability of new strong dynamics within the TC/ETC framework.

\section*{Acknowledgements}
This work was financially supported by the Academy of Finland project 267842.

\appendix

\section{Classification of the four-fermion operators from ETC}
\label{App:ETCclasses}
\addtocontents{toc}{\SkipTocEntry}
\subsubsection*{Class A: TC-singlet and QCD-singlet ETC bosons, with $Y=0$.}
At zero-momentum, the Lagrangian for type-A ETC bosons reads
\begin{eqnarray}
{\cal L}_{\rm ETC}^{\rm A} &=& \Big[g_{QQ}\, \overline{Q}_L \gamma_\mu Q_L + g_{qq}\, \overline{q}_L \gamma_\mu q_L  + g_{UU}\, \overline{U}_R \gamma_\mu U_R + g_{DD}\, \overline{D}_R \gamma_\mu D_R 
+ g_{tt}\, \overline{t}_R \gamma_\mu t_R + g_{bb}\, \overline{b}_R \gamma_\mu b_R\Big]A^\mu \nonumber \\
&+&\frac{1}{2}{\cal M}_A^2 A_\mu A^\mu\ ,
\end{eqnarray}
where TC and colour indices are suppressed, and all couplings are real. Integrating out the ETC field $A_\mu$ at tree-level gives
\begin{equation}
A_\mu=-\frac{1}{{\cal M}_A^2}\Big[g_{QQ}\, \overline{Q}_L \gamma_\mu Q_L + g_{qq}\, \overline{q}_L \gamma_\mu q_L  + g_{UU}\, \overline{U}_R \gamma_\mu U_R + g_{DD}\, \overline{D}_R \gamma_\mu D_R 
+ g_{tt}\, \overline{t}_R \gamma_\mu t_R + g_{bb}\, \overline{b}_R \gamma_\mu b_R\Big] \ ,
\end{equation}
whence, plugging back in ${\cal L}_{\rm ETC}^{\rm A}$,
{\allowdisplaybreaks\allowdisplaybreaks[4]
\begin{align}\allowdisplaybreaks
{\cal L}_{\rm ETC}^{\rm A} &= -\frac{1}{2}\frac{g_{QQ}^2}{{\cal M}_A^2}\left(\overline{Q}_L \gamma_\mu Q_L\right)^2
-\frac{1}{2}\frac{g_{qq}^2}{{\cal M}_A^2}\left(\overline{q}_L \gamma_\mu q_L\right)^2
-\frac{1}{2}\frac{g_{UU}^2}{{\cal M}_A^2}\left(\overline{U}_R \gamma_\mu U_R\right)^2
-\frac{1}{2}\frac{g_{DD}^2}{{\cal M}_A^2}\left(\overline{D}_R \gamma_\mu D_R\right)^2
-\frac{1}{2}\frac{g_{tt}^2}{{\cal M}_A^2}\left(\overline{t}_R \gamma_\mu t_R\right)^2 \nonumber \\
&-\frac{1}{2}\frac{g_{bb}^2}{{\cal M}_A^2}\left(\overline{b}_R \gamma_\mu b_R\right)^2
-\frac{g_{QQ}g_{qq}}{{\cal M}_A^2} \left(\overline{Q}_L \gamma_\mu Q_L\right)\left(\overline{q}_L \gamma^\mu q_L\right)
-\frac{g_{QQ}g_{UU}}{{\cal M}_A^2} \left(\overline{Q}_L \gamma_\mu Q_L\right)\left(\overline{U}_R \gamma^\mu U_R\right) \nonumber \\
&-\frac{g_{QQ}g_{DD}}{{\cal M}_A^2} \left(\overline{Q}_L \gamma_\mu Q_L\right)\left(\overline{D}_R \gamma^\mu D_R\right)
-\frac{g_{QQ}g_{tt}}{{\cal M}_A^2} \left(\overline{Q}_L \gamma_\mu Q_L\right)\left(\overline{t}_R \gamma^\mu t_R\right)
-\frac{g_{QQ}g_{bb}}{{\cal M}_A^2} \left(\overline{Q}_L \gamma_\mu Q_L\right)\left(\overline{b}_R \gamma^\mu b_R\right) \nonumber \\
&-\frac{g_{qq}g_{UU}}{{\cal M}_A^2} \left(\overline{q}_L \gamma_\mu q_L\right)\left(\overline{U}_R \gamma^\mu U_R\right)
-\frac{g_{qq}g_{DD}}{{\cal M}_A^2} \left(\overline{q}_L \gamma_\mu q_L\right)\left(\overline{D}_R \gamma^\mu D_R\right)
-\frac{g_{qq}g_{tt}}{{\cal M}_A^2} \left(\overline{q}_L \gamma_\mu q_L\right)\left(\overline{t}_R \gamma^\mu t_R\right) \nonumber \\
&-\frac{g_{qq}g_{bb}}{{\cal M}_A^2} \left(\overline{q}_L \gamma_\mu q_L\right)\left(\overline{b}_R \gamma^\mu b_R\right)
-\frac{g_{UU}g_{DD}}{{\cal M}_A^2} \left(\overline{U}_R \gamma_\mu U_R\right)\left(\overline{D}_R \gamma^\mu D_R\right)
-\frac{g_{UU}g_{tt}}{{\cal M}_A^2} \left(\overline{U}_R \gamma_\mu U_R\right)\left(\overline{t}_R \gamma^\mu t_R\right) \nonumber \\
&-\frac{g_{UU}g_{bb}}{{\cal M}_A^2} \left(\overline{U}_R \gamma_\mu U_R\right)\left(\overline{b}_R \gamma^\mu b_R\right)
-\frac{g_{DD}g_{tt}}{{\cal M}_A^2} \left(\overline{D}_R \gamma_\mu D_R\right)\left(\overline{t}_R \gamma^\mu t_R\right)
-\frac{g_{DD}g_{bb}}{{\cal M}_A^2} \left(\overline{D}_R \gamma_\mu D_R\right)\left(\overline{b}_R \gamma^\mu b_R\right) \nonumber \\
&-\frac{g_{tt}g_{bb}}{{\cal M}_A^2} \left(\overline{t}_R \gamma_\mu t_R\right)\left(\overline{b}_R \gamma^\mu b_R\right) \ .
\end{align}}
\addtocontents{toc}{\SkipTocEntry}
\subsubsection*{Class B: TC-singlet and QCD-singlet ETC bosons, with $Y=1$.}
At zero-momentum, the Lagrangian for type-B ETC bosons reads
\begin{eqnarray}
{\cal L}_{\rm ETC}^{\rm B} &=& \Big[g_{UD}\, \overline{U}_R \gamma_\mu D_R + g_{tb}\, \overline{t}_R \gamma_\mu b_R\Big]B^\mu + {\rm h.c.}
+{\cal M}_B^2 B_\mu^\ast B^\mu\ .
\end{eqnarray}
Integrating out the ETC field $B_\mu$ at tree-level gives
\begin{equation}
B_\mu^\ast = -\frac{1}{{\cal M}_B^2}\Big[g_{UD}\, \overline{U}_R \gamma_\mu D_R + g_{tb}\, \overline{t}_R \gamma_\mu b_R\Big] \ ,
\end{equation}
whence, plugging back in ${\cal L}_{\rm ETC}^{\rm B}$,
\begin{eqnarray}
{\cal L}_{\rm ETC}^{\rm B} &=& -\frac{|g_{UD}|^2}{{\cal M}_B^2} \left(\overline{U}_R \gamma_\mu D_R\right)\left(\overline{D}_R \gamma^\mu U_R\right)
-\frac{|g_{tb}|^2}{{\cal M}_B^2} \left(\overline{t}_R \gamma_\mu b_R\right)\left(\overline{b}_R \gamma^\mu t_R\right) \nonumber \\
&-&\left[ \frac{g_{UD}g_{tb}^\ast}{{\cal M}_B^2} \left(\overline{U}_R \gamma_\mu D_R\right)\left(\overline{b}_R \gamma^\mu t_R\right)+{\rm h.c.} \right]
\end{eqnarray}
\addtocontents{toc}{\SkipTocEntry}
\subsubsection*{Class C: TC-$N$ and QCD-$N_c$ ETC bosons, with $Y=1/6$.}
At zero-momentum, the Lagrangian for type-C ETC bosons reads
\begin{eqnarray}
{\cal L}_{\rm ETC}^{\rm C} &=& \Big[g_{Qq}\, \overline{Q}_L \gamma_\mu q_L + g_{Ut}\, \overline{U}_R \gamma_\mu t_R + g_{Db}\, \overline{D}_R \gamma_\mu b_R\Big]C^\mu + {\rm h.c.}
+{\cal M}_C^2 C_\mu^\ast C^\mu\ .
\end{eqnarray}
Integrating out the ETC field $C_\mu$ at tree-level gives
\begin{equation}
C_\mu^\ast=-\frac{1}{{\cal M}_C^2}\Big[g_{Qq}\, \overline{Q}_L \gamma_\mu q_L + g_{Ut}\, \overline{U}_R \gamma_\mu t_R + g_{Db}\, \overline{D}_R \gamma_\mu b_R\Big] \ ,
\end{equation}
whence, plugging back in ${\cal L}_{\rm ETC}^{\rm C}$,
\begin{eqnarray}
{\cal L}_{\rm ETC}^{\rm C} &=& -\frac{|g_{Qq}|^2}{{\cal M}_C^2}\left(\overline{Q}_L \gamma_\mu q_L\right)\left(\overline{q}_L \gamma^\mu Q_L\right)
-\frac{|g_{Ut}|^2}{{\cal M}_C^2}\left(\overline{U}_R \gamma_\mu t_R\right)\left(\overline{t}_R \gamma^\mu U_R\right)
-\frac{|g_{Db}|^2}{{\cal M}_C^2}\left(\overline{D}_R \gamma_\mu b_R\right)\left(\overline{b}_R \gamma^\mu D_R\right) \nonumber \\
&-&\left[\frac{g_{Qq}g_{Ut}^\ast}{{\cal M}_C^2}\left(\overline{Q}_L \gamma_\mu q_L\right)\left(\overline{t}_R \gamma^\mu U_R\right)+{\rm h.c.}\right]
-\left[\frac{g_{Qq}g_{Db}^\ast}{{\cal M}_C^2}\left(\overline{Q}_L \gamma_\mu q_L\right)\left(\overline{b}_R \gamma^\mu D_R\right)+{\rm h.c.}\right] \nonumber \\
&-&\left[\frac{g_{Ut}g_{Db}^\ast}{{\cal M}_C^2}\left(\overline{U}_R \gamma_\mu t_R\right)\left(\overline{b}_R \gamma^\mu D_R\right)+{\rm h.c.}\right]\ .
\end{eqnarray}
\addtocontents{toc}{\SkipTocEntry}
\subsubsection*{Class D: TC-$N$ and QCD-$N_c$ ETC bosons, with $Y=5/6$.}
At zero-momentum, the Lagrangian for type-D ETC bosons reads
\begin{eqnarray}
{\cal L}_{\rm ETC}^{\rm D} &=& g_{Ub}\, \overline{U}_R \gamma_\mu b_R D^\mu+ {\rm h.c.}
+{\cal M}_D^2 D_\mu^\ast D^\mu\ .
\end{eqnarray}
Integrating out the ETC field $D_\mu$ at tree-level gives
\begin{equation}
D_\mu^\ast=-\frac{1}{{\cal M}_D^2}g_{Ub}\, \overline{U}_R \gamma_\mu b_R \ ,
\end{equation}
whence, plugging back in ${\cal L}_{\rm ETC}^{\rm D}$,
\begin{eqnarray}
{\cal L}_{\rm ETC}^{\rm D} &=& -\frac{|g_{Ub}|^2}{{\cal M}_D^2}\left(\overline{U}_R \gamma_\mu b_R\right)\left(\overline{b}_R \gamma^\mu U_R\right)\ .
\end{eqnarray}
\addtocontents{toc}{\SkipTocEntry}
\subsubsection*{Class E: TC-$N$ and QCD-$N_c$ ETC bosons, with $Y=7/6$.}
At zero-momentum, the Lagrangian for type-E ETC bosons reads
\begin{eqnarray}
{\cal L}_{\rm ETC}^{\rm E} &=& g_{Dt}\, \overline{D}_R \gamma_\mu t_R E^\mu+ {\rm h.c.}
+{\cal M}_E^2 E_\mu^\ast E^\mu\ .
\end{eqnarray}
Integrating out the ETC field $E_\mu$ at tree-level gives
\begin{equation}
E_\mu^\ast =-\frac{1}{{\cal M}_E^2}g_{Dt}\, \overline{D}_R \gamma_\mu t_R \ ,
\end{equation}
whence, plugging back in ${\cal L}_{\rm ETC}^{\rm E}$,
\begin{eqnarray}
{\cal L}_{\rm ETC}^{\rm E} &=& -\frac{|g_{Dt}|^2}{{\cal M}_E^2}\left(\overline{D}_R \gamma_\mu t_R\right)\left(\overline{t}_R \gamma^\mu D_R\right)\ .
\end{eqnarray}

\section{Fierz rearrangement formulas}
\label{App:Fierz}
We use the following Fierz rearrangement formulas for anticommuting fields to simplify the products: 
{\allowdisplaybreaks\allowdisplaybreaks[4]
\begin{align}\allowdisplaybreaks
&-\left(\overline{Q}_L \gamma_\mu Q_L\right)\left(\overline{U}_R \gamma^\mu U_R\right)=\frac{2}{N} \left(\overline{Q}_L  U_R\right)\left(\overline{U}_R Q_L\right)
+4 \left(\overline{Q}_L T_{\rm TC}^A U_R\right)\left(\overline{U}_R T_{\rm TC}^A Q_L\right)\ , \nonumber \\
&-\left(\overline{Q}_L \gamma_\mu Q_L\right)\left(\overline{D}_R \gamma^\mu D_R\right)=\frac{2}{N} \left(\overline{Q}_L  D_R\right)\left(\overline{D}_R Q_L\right)
+4 \left(\overline{Q}_L T_{\rm TC}^A D_R\right)\left(\overline{D}_R T_{\rm TC}^A Q_L\right)\ , \nonumber \\
&-\left(\overline{q}_L \gamma_\mu q_L\right)\left(\overline{t}_R \gamma^\mu t_R\right)=\frac{2}{N_c} \left(\overline{q}_L  t_R\right)\left(\overline{t}_R q_L\right)
+4 \left(\overline{q}_L T_{\rm QCD}^a t_R\right)\left(\overline{t}_R T_{\rm QCD}^a q_L\right)\ , \nonumber \\
&-\left(\overline{q}_L \gamma_\mu q_L\right)\left(\overline{b}_R \gamma^\mu b_R\right)=\frac{2}{N_c} \left(\overline{q}_L  b_R\right)\left(\overline{b}_R q_L\right)
+4 \left(\overline{q}_L T_{\rm QCD}^a b_R\right)\left(\overline{b}_R T_{\rm QCD}^a q_L\right)\ , \nonumber \\
&-\left(\overline{Q}_L \gamma_\mu q_L\right)\left(\overline{t}_R \gamma^\mu U_R\right)=2\left(\overline{Q}_L  U_R\right)\left(\overline{t}_R  q_L\right)\ , \nonumber \\
&-\left(\overline{Q}_L \gamma_\mu q_L\right)\left(\overline{b}_R \gamma^\mu D_R\right)=2\left(\overline{Q}_L  D_R\right)\left(\overline{b}_R  q_L\right)\ ,  \nonumber \\
&-\left(\overline{Q}_L \gamma_\mu q_L\right)\left(\overline{q}_L \gamma^\mu Q_L\right) = -\frac{1}{2}\left(\overline{Q}_L \gamma_\mu Q_L\right)\left(\overline{q}_L \gamma^\mu q_L\right) 
-2\left(\overline{Q}_L \gamma_\mu T^i Q_L\right)\left(\overline{q}_L \gamma^\mu T^i q_L\right)\ , \nonumber \\
&-\left(\overline{t}_R \gamma_\mu U_R\right)\left(\overline{U}_R \gamma^\mu t_R\right)=-\left(\overline{U}_R \gamma_\mu U_R\right)\left(\overline{t}_R \gamma^\mu t_R\right)\ ,  \nonumber \\
&-\left(\overline{b}_R \gamma_\mu U_R\right)\left(\overline{U}_R \gamma^\mu b_R\right)=-\left(\overline{U}_R \gamma_\mu U_R\right)\left(\overline{b}_R \gamma^\mu b_R\right)\ ,  \nonumber \\
&-\left(\overline{b}_R \gamma_\mu D_R\right)\left(\overline{D}_R \gamma^\mu b_R\right)=-\left(\overline{D}_R \gamma_\mu D_R\right)\left(\overline{b}_R \gamma^\mu b_R\right)\ ,  \nonumber \\
&-\left(\overline{t}_R \gamma_\mu D_R\right)\left(\overline{D}_R \gamma^\mu t_R\right)=-\left(\overline{D}_R \gamma_\mu D_R\right)\left(\overline{t}_R \gamma^\mu t_R\right)\ ,  \nonumber \\
&-\left(\overline{U}_R \gamma_\mu t_R\right)\left(\overline{b}_R \gamma^\mu D_R\right)=-\left(\overline{U}_R \gamma_\mu D_R\right)\left(\overline{b}_R \gamma^\mu t_R\right) \ .
\end{align}}
%
%
%
%
%
%
%
%
%
%
%
%
%
%
%
%
%
%
%
%
%
%
%
%
\section{Integrals}\label{Sec:Int}
We provide expressions for the standard integrals shown at the end of Sec. \ref{Sec:Cutoff}. For definiteness, we use the cutoff $\Lambda$.
{\allowdisplaybreaks\allowdisplaybreaks[4]
\begin{align}\allowdisplaybreaks
I_X & = \frac{1}{16\pi^2}\left[\Lambda^2-M_X^2\, \log\, \frac{\Lambda^2+M_X^2}{M_X^2}\right] \ , \\
J_{XY} & = \frac{1}{16\pi^2}\, \left[\frac{M_X^2}{M_X^2-M_Y^2}\, \log\, \frac{\Lambda^2+M_X^2}{M_X^2}
-\frac{M_Y^2}{M_X^2-M_Y^2}\, \log\, \frac{\Lambda^2+M_Y^2}{M_Y^2}\right] \\
K_{XY} & = \frac{1}{32\pi^2}\left[
\frac{M_X^2\left(M_X^2-2\, M_Y^2\right)}{\left(M_X^2-M_Y^2\right)^2}\, \log\frac{\Lambda^2+M_X^2}{M_X^2}
+\frac{M_Y^4}{\left(M_X^2-M_Y^2\right)^2}\, \log\frac{\Lambda^2+M_Y^2}{M_Y^2}\right. \nonumber \\
&\left. \quad\quad\quad\ 
+\frac{\Lambda^4}{\left(M_X^2-M_Y^2\right)^2}\, \log\frac{\Lambda^2+M_X^2}{\Lambda^2+M_Y^2}
-\frac{\Lambda^2}{M_X^2-M_Y^2}\right] \ .
\end{align}
The expression for $L_{XY}$ is rather complicated for $X\neq Y$. However for our computations we only need $L_{XX}$, which is much simpler. We have
\begin{eqnarray}
J_{XX} = 2\, K_{XX} = 6\, L_{XX} = \frac{1}{16\pi^2}\left[
\log\frac{\Lambda^2+M_X^2}{M_X^2}-\frac{\Lambda^2}{\Lambda^2+M_X^2}\right]
\label{Eq:RelInt}
\end{eqnarray}
Finally, we have
\begin{eqnarray}
J_{XX}^\prime = \frac{1}{96\pi^2}\, \frac{1}{\left(1+M_X^2/\Lambda^2\right)^2} \ .
\end{eqnarray}

\section{Two-loop contributions to $S$ and $T$ parameters}
\label{app:twoloopST}
\begin{eqnarray}
S_{QQQQ} &=& -16\pi\frac{g_{QQ}^2}{{\cal M}^2}\, N^2\, \left[\frac{J_{DD}-J_{UU}}{3}+2\left(J_{UU}^\prime-J_{DD}^\prime\right)\right] \left(J_{UU}\, M_U^2-J_{DD}\, M_D^2\right)\nonumber \\
&=&-S_0\, \frac{g_{QQ}^2\, N^2}{8\pi^2}\left(\frac{M_U^2}{{\cal M}^2}\, {\overline \log}\frac{\Lambda^2}{M_U^2}-\frac{M_D^2}{{\cal M}^2}\, {\overline \log}\frac{\Lambda^2}{M_D^2}\right)
\log \frac{M_U^2}{M_D^2} 
\left[1+{\cal O}\left(\frac{M_Q^2}{\Lambda^2}\right)+{\cal O}\left(\frac{M_t^2}{{\cal M}^2}\right)\right]
\end{eqnarray}
\begin{eqnarray}
T_{QQQQ} &=& \frac{4\sqrt2\, G_F}{\alpha}\,  \frac{g_{QQ}^2\, N^2}{{\cal M}^2}\, \left(J_{UU}\, M_U^2-J_{DD}\, M_D^2\right)^2 \nonumber \\
&=& T_0\, \frac{g_{QQ}^2\, N^2}{4\pi^2}\left(\frac{M_U^2}{{\cal M}\, M_t}{\overline \log}\frac{\Lambda^2}{M_U^2}-\frac{M_D^2}{{\cal M}\, M_t}{\overline \log}\frac{\Lambda^2}{M_D^2}\right)^2
\left[1+{\cal O}\left(\frac{M_Q^2}{\Lambda^2}\right)+{\cal O}\left(\frac{M_t^2}{{\cal M}^2}\right)\right]
\end{eqnarray}

\begin{eqnarray}
S_{UUUU} &=& 16\pi\frac{g_{UU}^2}{{\cal M}^2}\, N^2\, \left(\frac{1}{3}J_{UU}-2J_{UU}^\prime\right)J_{UU}\, M_U^2 \nonumber \\
&=& S_0\, \frac{g_{UU}^2\, N^2}{8\pi^2}\, \frac{M_U^2}{{\cal M}^2}\, {\overline \log}\frac{\Lambda^2}{M_U^2}\left({\overline \log}\frac{\Lambda^2}{M_U^2}-1\right)
\left[1+{\cal O}\left(\frac{M_Q^2}{\Lambda^2}\right)+{\cal O}\left(\frac{M_t^2}{{\cal M}^2}\right)\right]
\end{eqnarray}
\begin{eqnarray}
T_{UUUU} &=& \frac{4\sqrt2\, G_F}{\alpha}\,  \frac{g_{UU}^2\, N^2}{{\cal M}^2}\, \left(J_{UU}\, M_U^2\right)^2 \nonumber \\
&=& T_0\, \frac{g_{UU}^2\, N^2}{4\pi^2}\left(\frac{M_U^2}{{\cal M}\, M_t}{\overline \log}\frac{\Lambda^2}{M_U^2}\right)^2
\left[1+{\cal O}\left(\frac{M_Q^2}{\Lambda^2}\right)+{\cal O}\left(\frac{M_t^2}{{\cal M}^2}\right)\right]
\label{Eq:TUUUU}
\end{eqnarray}

\begin{eqnarray}
S_{DDDD} &=& 16\pi\frac{g_{DD}^2}{{\cal M}^2}\, N^2\, \left(\frac{1}{3}J_{DD}-2J_{DD}^\prime\right)J_{DD}\, M_D^2 \nonumber \\
&=& S_0\, \frac{g_{DD}^2\, N^2}{8\pi^2}\, \frac{M_D^2}{{\cal M}^2}\, {\overline \log}\frac{\Lambda^2}{M_D^2}\left({\overline \log}\frac{\Lambda^2}{M_D^2}-1\right)
\left[1+{\cal O}\left(\frac{M_Q^2}{\Lambda^2}\right)+{\cal O}\left(\frac{M_t^2}{{\cal M}^2}\right)\right]
\end{eqnarray}
\begin{eqnarray}
T_{DDDD} &=& \frac{4\sqrt2\, G_F}{\alpha}\,  \frac{g_{DD}^2\, N^2}{{\cal M}^2}\, \left(J_{DD}\, M_D^2\right)^2 \nonumber \\
&=& T_0\, \frac{g_{DD}^2\, N^2}{4\pi^2}\left(\frac{M_D^2}{{\cal M}\, M_t}{\overline \log}\frac{\Lambda^2}{M_D^2}\right)^2
\left[1+{\cal O}\left(\frac{M_Q^2}{\Lambda^2}\right)+{\cal O}\left(\frac{M_t^2}{{\cal M}^2}\right)\right]
\label{Eq:TDDDD}
\end{eqnarray}

\begin{eqnarray}
S_{UUDD} &=& -16\pi\, \frac{g_{UU}g_{DD}}{{\cal M}^2}\, N^2\, \left(
J_{UU}\, J_{DD}\frac{M_U^2+M_D^2}{3}-2\, J_{UU}\, J_{DD}^\prime\, M_U^2-2\, J_{DD}\, J_{UU}^\prime\, M_D^2\right) \nonumber \\
&=&-S_0\, \frac{g_{UU}g_{DD}\, N^2}{8\pi^2}\left(
\frac{M_U^2+M_D^2}{{\cal M}^2}{\overline \log}\frac{\Lambda^2}{M_U^2} {\overline \log}\frac{\Lambda^2}{M_D^2}
-\frac{M_U^2}{{\cal M}^2} {\overline \log}\frac{\Lambda^2}{M_U^2}
-\frac{M_D^2}{{\cal M}^2} {\overline \log}\frac{\Lambda^2}{M_D^2}\right)\nonumber \\
&&\times \left[1+{\cal O}\left(\frac{M_Q^2}{\Lambda^2}\right)+{\cal O}\left(\frac{M_t^2}{{\cal M}^2}\right)\right]
\end{eqnarray}
\begin{equation}
S_{UDUD}=0
\end{equation}
\begin{eqnarray}
T_{UUDD} &=& -\frac{4\sqrt2\, G_F}{\alpha}\frac{2g_{UU}g_{DD}}{{\cal M}^2}\, N^2\, J_{UU}\, J_{DD}\, M_U^2\, M_D^2 \nonumber \\
&=& -T_0\, \frac{g_{UU}g_{DD}\, N^2}{2\pi^2}\frac{M_U^2\, M_D^2}{{\cal M}^2\, M_t^2}
{\overline\log}\frac{\Lambda^2}{M_U^2}{\overline\log}\frac{\Lambda^2}{M_D^2}\left[1+{\cal O}\left(\frac{M_Q^2}{\Lambda^2}\right)+{\cal O}\left(\frac{M_t^2}{{\cal M}^2}\right)\right]
\label{Eq:TUUDD}
\end{eqnarray}
\begin{eqnarray}
T_{UDUD} &=& -\frac{4\sqrt2\, G_F}{\alpha}\frac{2g_{UD}^2}{{\cal M}^2}\, N^2\, J^2_{UD}M_U^2\, M_D^2 \nonumber \\
&=&-T_0 \frac{g_{UD}^2\, N^2}{2\pi^2}\, \frac{M_U^2\, M_D^2}{{\cal M}^2\, M_t^2}\, 
\left[\frac{M_U^2 {\overline \log}(\Lambda^2/M_U^2)-M_D^2 {\overline \log}(\Lambda^2/M_D^2)}{M_U^2-M_D^2}+1\right]^2 \nonumber \\
&\times&\left[1+{\cal O}\left(\frac{M_Q^2}{\Lambda^2}\right)+{\cal O}\left(\frac{M_t^2}{{\cal M}^2}\right)\right]
\label{Eq:UDtparameter}
\end{eqnarray}

\begin{eqnarray}
S_{QQqq} &=& -16\pi\, \frac{g_{QQ}g_{qq}}{{\cal M}^2}\, N\, N_c\, \Bigg[
\left(\frac{J_{DD}-J_{UU}}{3}+2\, J_{UU}^\prime-2\, J_{DD}^\prime\right)J_{tt}\, M_t^2\nonumber \\
&&+\left(\frac{4\, J_{bb}-2\, J_{tt}}{9}+2\, J_{tt}^\prime-2\, J_{bb}^\prime\right)(J_{UU}M^2_{U}-J_{DD} M_D^2)\Bigg] \nonumber \\
&=&-S_0\, \frac{g_{QQ}g_{qq}\, N\, N_c}{8\pi^2}
\left[\left({\overline \log}\frac{\Lambda^2}{M_D^2}-{\overline \log}\frac{\Lambda^2}{M_U^2}\right)\, {\overline \log}\frac{{\cal M}^2}{M_t^2}\, \frac{M_t^2}{{\cal M}^2}
+\frac{4\, {\overline \log}({\cal M}^2/M_b^2)-2\, {\overline \log}({\cal M}^2/M_t^2)}{3} \right. \nonumber \\
&&\quad\quad\quad\quad\quad\quad\quad\ \  \left.\times\left(\frac{M_U^2}{{\cal M}^2}\, {\overline \log}\frac{\Lambda^2}{M_U^2}-\frac{M_D^2}{{\cal M}^2}\, {\overline \log}\frac{\Lambda^2}{M_D^2}\right)\right]
\left[1+{\cal O}\left(\frac{M_Q^2}{\Lambda^2}\right)+{\cal O}\left(\frac{M_t^2}{{\cal M}^2}\right)\right]
\end{eqnarray}
\begin{eqnarray}
S_{QqQq} &=& 16\pi\, \frac{g_{Qq}^2}{2{\cal M}^2}\, N\, N_c \Bigg[
\left(\frac{2\, J_{UU}}{3}-4\, J_{UU}^\prime\right)J_{tt}\, M_t^2
+\left(\frac{4\, J_{bb}+2\, J_{tt}}{9}-2\, J_{tt}^\prime-2\, J_{bb}^\prime\right)\left(J_{UU}\, M_U^2+J_{DD}\, M_D^2\right) \nonumber \\
&&\quad\quad\quad\quad\quad\quad\ \ -\left(\frac{4\, J_{bb}-2\, J_{tt}}{9}+2\, J_{tt}^\prime-2\, J_{bb}^\prime\right)\left(J_{UU}\, M_U^2-J_{DD}\, M_D^2\right)\Bigg] \nonumber \\
&=&S_0\, \frac{g_{Qq}^2\, N\, N_c}{8\pi^2} \left[
\left({\overline \log}\frac{\Lambda^2}{M_U^2}-1\right){\overline \log}\frac{{\cal M}^2}{M_t^2}\, \frac{M_t^2}{{\cal M}^2}
+\left(\frac{2\, {\overline \log}({\cal M}^2/M_b^2) + {\overline \log}({\cal M}^2/M_t^2)}{3}-1\right)\right. \nonumber \\
&\times& \left. \left(\frac{M_U^2}{{\cal M}^2}\, {\overline \log}\frac{\Lambda^2}{M_U^2}+\frac{M_D^2}{{\cal M}^2}\, {\overline \log}\frac{\Lambda^2}{M_D^2}\right)
-\frac{2\, {\overline \log}({\cal M}^2/M_b^2) - {\overline \log}({\cal M}^2/M_t^2)}{3}
\left(\frac{M_U^2}{{\cal M}^2}\, {\overline \log}\frac{\Lambda^2}{M_U^2}-\frac{M_D^2}{{\cal M}^2}\, {\overline \log}\frac{\Lambda^2}{M_D^2}\right)\right] \nonumber \\
&\times& \left[1+{\cal O}\left(\frac{M_Q^2}{\Lambda^2}\right)+{\cal O}\left(\frac{M_t^2}{{\cal M}^2}\right)\right] 
\end{eqnarray}
\begin{eqnarray}
T_{QQqq} &=& \frac{4\sqrt2\, G_F}{\alpha}\, \frac{2g_{QQ}g_{qq}}{{\cal M}^2}\, N\, N_c\, \left(J_{UU}\, M_U^2-J_{DD}\, M_D^2\right)J_{tt}\, M_t^2 \nonumber \\
&=&T_0\, \frac{g_{QQ}g_{qq}\, N\, N_c}{2\pi^2}
\left(\frac{M_U^2}{{\cal M}^2}\, {\overline \log}\frac{\Lambda^2}{M_U^2}-\frac{M_D^2}{{\cal M}^2}\, {\overline \log}\frac{\Lambda^2}{M_D^2}\right){\overline \log}\frac{{\cal M}^2}{M_t^2}
\left[1+{\cal O}\left(\frac{M_Q^2}{\Lambda^2}\right)+{\cal O}\left(\frac{M_t^2}{{\cal M}^2}\right)\right]
\end{eqnarray}
\begin{eqnarray}
T_{QqQq} &=& \frac{4\sqrt2\, G_F}{\alpha}\, \frac{g_{Qq}^2}{{\cal M}^2}\, N\, N_c
\left[2\, J_{UU}\, J_{tt}\, M_U^2-4\left(K_{UD}\, M_U^2+K_{DU}\, M_D^2\right)K_{tb}\right]M_t^2 \nonumber \\
&=& T_0\, \frac{g_{Qq}^2\, N\, N_c}{8\pi^2}\left[
2\left(\frac{M_U^2}{{\cal M}^2}\, {\overline \log}\frac{\Lambda^2}{M_U^2}-\frac{M_D^2}{{\cal M}^2}\, {\overline \log}\frac{\Lambda^2}{M_D^2}\right){\overline \log}\frac{{\cal M}^2}{M_t^2}
-\left({\overline \log}\frac{{\cal M}^2}{M_t^2}+\frac{1}{2}\right)\frac{M_U^2+M_D^2}{{\cal M}^2}\right. \nonumber \\
&-&\left.
\frac{M_U^4\, {\overline \log}(\Lambda^2/M_U^2)-M_D^4\, {\overline \log}(\Lambda^2/M_D^2)
-2\log (M_U^2/M_D^2){\overline \log}({\cal M}^2/M_t^2)\, M_U^2\, M_D^2}{\left(M_U^2-M_D^2\right){\cal M}^2}\right] \nonumber \\
&\times& \left[1+{\cal O}\left(\frac{M_Q^2}{\Lambda^2}\right)+{\cal O}\left(\frac{M_t^2}{{\cal M}^2}\right)\right]
\end{eqnarray}

\begin{eqnarray}
S_{QQtt} &=& -16\pi\, \frac{g_{QQ}g_{tt}}{{\cal M}^2}\, N\, N_c \left[
\left(J_{UU}\, M_U^2-J_{DD}\, M_D^2\right)\left(\frac{4}{9}\, J_{tt}-2\, J_{tt}^\prime\right)
-\left(\frac{J_{DD}-J_{UU}}{3}+2\, J_{UU}^\prime-2\, J_{DD}^\prime\right)J_{tt}\, M_t^2\right] \nonumber \\
&=& -S_0\, \frac{g_{QQ}g_{tt}\, N\, N_c}{8\pi^2}\left[
\left(\frac{4}{3}\, {\overline \log}\frac{{\cal M}^2}{M_t^2}-1\right)\left(\frac{M_U^2}{{\cal M}^2}\, {\overline \log}\frac{\Lambda^2}{M_U^2}-\frac{M_D^2}{{\cal M}^2}\, {\overline \log}\frac{\Lambda^2}{M_D^2}\right)
-\frac{M_t^2}{{\cal M}^2}\, \log\frac{M_U^2}{M_D^2}\, {\overline \log}\frac{{\cal M}^2}{M_t^2}\right] \nonumber \\
&\times&\left[1+{\cal O}\left(\frac{M_Q^2}{\Lambda^2}\right)+{\cal O}\left(\frac{M_t^2}{{\cal M}^2}\right)\right]
\end{eqnarray}
\begin{eqnarray}
T_{QQtt} &=& - \frac{4\sqrt2\, G_F}{\alpha}\, \frac{2\, g_{QQ}g_{tt}}{{\cal M}^2}\, N\, N_c \left(J_{UU}\, M_U^2-J_{DD}\, M_D^2\right)J_{tt}\, M_t^2 \nonumber \\
&=&-T_0\, \frac{g_{QQ}g_{tt}\, N\, N_c}{2\pi^2}
\left(\frac{M_U^2}{{\cal M}^2}\, {\overline \log}\frac{\Lambda^2}{M_U^2}-\frac{M_D^2}{{\cal M}^2}\, {\overline \log}\frac{\Lambda^2}{M_D^2}\right)\, {\overline \log}\frac{{\cal M}^2}{M_t^2}
\left[1+{\cal O}\left(\frac{M_Q^2}{\Lambda^2}\right)+{\cal O}\left(\frac{M_t^2}{{\cal M}^2}\right)\right]
\end{eqnarray}

\begin{eqnarray}
S_{QQbb} &=& 16\pi\, \frac{g_{QQ}g_{bb}}{{\cal M}^2}\, N\, N_c \left(\frac{2}{9}\, J_{bb}-2\, J_{bb}^\prime\right)\left(J_{UU}\, M_U^2-J_{DD}\, M_D^2\right) \nonumber \\
&=& S_0\, \frac{g_{QQ}g_{bb}\, N\, N_c}{8\pi^2}
\left(\frac{M_U^2}{{\cal M}^2}\, {\overline \log}\frac{\Lambda^2}{M_U^2}-\frac{M_D^2}{{\cal M}^2}\, {\overline \log}\frac{\Lambda^2}{M_D^2}\right)
\left(\frac{2}{3}\, {\overline \log}\frac{{\cal M}^2}{M_b^2}-1\right) \nonumber \\
&\times&\left[1+{\cal O}\left(\frac{M_Q^2}{\Lambda^2}\right)+{\cal O}\left(\frac{M_t^2}{{\cal M}^2}\right)\right]
\end{eqnarray}
\begin{equation}
T_{QQbb} = 0
\end{equation}

\begin{eqnarray}
S_{qqUU} &=& 16\pi\, \frac{g_{qq}g_{UU}}{{\cal M}^2}\, N\, N_c
\left[\left(\frac{4\, J_{bb}-2\, J_{tt}}{9}+2\, J^\prime_{tt}-2\, J^\prime_{bb}\right)J_{UU}\, M_U^2
-\left(\frac{1}{3}\, J_{UU}-2J_{UU}^\prime\right)J_{tt}\, M_t^2\right] \nonumber \\
&=&S_0\, \frac{g_{qq}g_{UU}\, N\, N_c}{8\pi^2}\left[
\frac{M_U^2}{{\cal M}^2}\, \frac{4\, {\overline \log}({\cal M}^2/M_b^2)-2\, {\overline \log}({\cal M}^2/M_t^2)}{3}\, {\overline \log}\frac{\Lambda^2}{M_U^2}
-\frac{M_t^2}{{\cal M}^2}\, \left({\overline \log}\frac{\Lambda^2}{M_U^2}-1\right)\, {\overline \log}\frac{{\cal M}^2}{M_t^2}\right] \nonumber \\
&\times& \left[1+{\cal O}\left(\frac{M_Q^2}{\Lambda^2}\right)+{\cal O}\left(\frac{M_t^2}{{\cal M}^2}\right)\right]
\end{eqnarray}
\begin{eqnarray}
T_{qqUU} &=& -\frac{4\sqrt2\, G_F}{\alpha}\, \frac{2g_{qq}g_{UU}}{{\cal M}^2}\, N\, N_c\, J_{UU}\, J_{tt}\, M_U^2\, M_t^2 \nonumber \\
&=&-T_0\, \frac{g_{qq}g_{UU}\, N\, N_c}{2\pi^2}\, \frac{M_U^2}{{\cal M}^2}\, {\overline \log}\frac{\Lambda^2}{M_U^2}\, {\overline \log}\frac{{\cal M}^2}{M_t^2}
\left[1+{\cal O}\left(\frac{M_Q^2}{\Lambda^2}\right)+{\cal O}\left(\frac{M_t^2}{{\cal M}^2}\right)\right]
\end{eqnarray}

\begin{eqnarray}
S_{qqDD} &=& -16\pi\, \frac{g_{qq}g_{DD}}{{\cal M}^2}\, N\, N_c
\left[\left(\frac{4\, J_{bb}-2\, J_{tt}}{9}+2\, J^\prime_{tt}-2\, J^\prime_{bb}\right)J_{DD}\, M_D^2
-\left(\frac{1}{3}\, J_{DD}-2J_{DD}^\prime\right)J_{tt}\, M_t^2\right] \nonumber \\
&=&-S_0\, \frac{g_{qq}g_{DD}\, N\, N_c}{8\pi^2}\left[
\frac{M_D^2}{{\cal M}^2}\, \frac{4\, {\overline \log}({\cal M}^2/M_b^2)-2\, {\overline \log}({\cal M}^2/M_t^2)}{3}\, {\overline \log}\frac{\Lambda^2}{M_D^2}
-\frac{M_t^2}{{\cal M}^2}\, \left({\overline \log}\frac{\Lambda^2}{M_D^2}-1\right)\, {\overline \log}\frac{{\cal M}^2}{M_t^2}\right] \nonumber \\
&\times& \left[1+{\cal O}\left(\frac{M_Q^2}{\Lambda^2}\right)+{\cal O}\left(\frac{M_t^2}{{\cal M}^2}\right)\right]
\end{eqnarray}
\begin{eqnarray}
T_{qqDD} &=& \frac{4\sqrt2\, G_F}{\alpha}\, \frac{2g_{qq}g_{DD}}{{\cal M}^2}\, N\, N_c\, J_{DD}\, J_{tt}\, M_D^2\, M_t^2 \nonumber \\
&=&T_0\, \frac{g_{qq}g_{DD}\, N\, N_c}{2\pi^2}\, \frac{M_D^2}{{\cal M}^2}\, {\overline \log}\frac{\Lambda^2}{M_D^2}\, {\overline \log}\frac{{\cal M}^2}{M_t^2}
\left[1+{\cal O}\left(\frac{M_Q^2}{\Lambda^2}\right)+{\cal O}\left(\frac{M_t^2}{{\cal M}^2}\right)\right]
\end{eqnarray}

\begin{eqnarray}
S_{UUtt}+S_{UtUt} &=& 16\pi\, \frac{\left(g_{UU}g_{tt}+g_{Ut}^2\right)}{{\cal M}^2}\, N\, N_c
\left[\left(\frac{4}{9}\, J_{tt}-2\, J_{tt}^\prime\right)J_{UU}\, M_U^2
+\left(\frac{1}{3}\, J_{UU}-2\, J_{UU}^\prime\right)J_{tt}\, M_t^2\right] \nonumber \\
&=&S_0\, \frac{\left(g_{UU}g_{tt}+g_{Ut}^2\right)\, N\, N_c}{8\pi^2}
\left[ \frac{M_U^2}{{\cal M}^2}\, \left(\frac{4}{3}\, {\overline \log}\frac{{\cal M}^2}{M_t^2}-1\right){\overline \log}\frac{\Lambda^2}{M_U^2}
+\frac{M_t^2}{{\cal M}^2}\, \left({\overline \log}\frac{\Lambda^2}{M_U^2}-1\right){\overline \log}\frac{{\cal M}^2}{M_t^2}\right] \nonumber \\
&\times& \left[1+{\cal O}\left(\frac{M_Q^2}{\Lambda^2}\right)+{\cal O}\left(\frac{M_t^2}{{\cal M}^2}\right)\right]
\end{eqnarray}
\begin{eqnarray}
T_{UUtt}+T_{UtUt} &=& \frac{4\sqrt2\, G_F}{\alpha}\, \frac{2\left(g_{UU}g_{tt}+g_{Ut}^2\right)}{{\cal M}^2}\, N\, N_c\, J_{UU}\, J_{tt}\, M_U^2\, M_t^2 \nonumber \\
&=& T_0\, \frac{\left(g_{UU}g_{tt}+g_{Ut}^2\right)\, N\, N_c}{2\pi^2}
\frac{M_U^2}{{\cal M}^2}\, {\overline \log}\frac{\Lambda^2}{M_U^2}\, {\overline \log}\frac{{\cal M}^2}{M_t^2}
\left[1+{\cal O}\left(\frac{M_Q^2}{\Lambda^2}\right)+{\cal O}\left(\frac{M_t^2}{{\cal M}^2}\right)\right]
\label{Eq:TUtUt}
\end{eqnarray}

\begin{eqnarray}
S_{DDtt}+S_{DtDt} &=& -16\pi\, \frac{\left(g_{DD}g_{tt}+g_{Dt}^2\right)}{{\cal M}^2}\, N\, N_c
\left[\left(\frac{4}{9}\, J_{tt}-2\, J_{tt}^\prime\right)J_{DD}\, M_D^2
+\left(\frac{1}{3}\, J_{DD}-2\, J_{DD}^\prime\right)J_{tt}\, M_t^2\right] \nonumber \\
&=&-S_0\, \frac{\left(g_{DD}g_{tt}+g_{Dt}^2\right)\, N\, N_c}{8\pi^2}
\left[ \frac{M_D^2}{{\cal M}^2}\, \left(\frac{4}{3}\, {\overline \log}\frac{{\cal M}^2}{M_t^2}-1\right){\overline \log}\frac{\Lambda^2}{M_D^2}
+\frac{M_t^2}{{\cal M}^2}\, \left({\overline \log}\frac{\Lambda^2}{M_D^2}-1\right){\overline \log}\frac{{\cal M}^2}{M_t^2}\right] \nonumber \\
&\times& \left[1+{\cal O}\left(\frac{M_Q^2}{\Lambda^2}\right)+{\cal O}\left(\frac{M_t^2}{{\cal M}^2}\right)\right]
\label{Eq:SDtDt}
\end{eqnarray}
\begin{eqnarray}
T_{DDtt}+T_{DtDt} &=& -\frac{4\sqrt2\, G_F}{\alpha}\, \frac{2\left(g_{DD}g_{tt}+g_{Dt}^2\right)}{{\cal M}^2}\, N\, N_c\, J_{DD}\, J_{tt}\, M_D^2\, M_t^2 \nonumber \\
&=& - T_0\, \frac{\left(g_{DD}g_{tt}+g_{Dt}^2\right)\, N\, N_c}{2\pi^2}
\frac{M_D^2}{{\cal M}^2}\, {\overline \log}\frac{\Lambda^2}{M_D^2}\, {\overline \log}\frac{{\cal M}^2}{M_t^2}
\left[1+{\cal O}\left(\frac{M_Q^2}{\Lambda^2}\right)+{\cal O}\left(\frac{M_t^2}{{\cal M}^2}\right)\right]
\label{Eq:TDtDt}
\end{eqnarray}

\begin{eqnarray}
S_{UUbb}+S_{UbUb} &=& -16\pi\, \frac{g_{UU}g_{bb}+g_{Ub}^2}{{\cal M}^2}\, N\, N_c \left(\frac{2}{9}\, J_{bb}-2\, J_{bb}^\prime\right) J_{UU}\, M_U^2 \nonumber \\
&=& -S_0\, \frac{\left(g_{UU}g_{bb}+g_{Ub}^2\right)\, N\, N_c}{8\pi^2}
\frac{M_U^2}{{\cal M}^2}\, \left(\frac{2}{3}\, {\overline \log}\frac{{\cal M}^2}{M_b^2}-1\right){\overline \log}\frac{\Lambda^2}{M_U^2} \nonumber \\
&\times&\left[1+{\cal O}\left(\frac{M_Q^2}{\Lambda^2}\right)+{\cal O}\left(\frac{M_t^2}{{\cal M}^2}\right)\right]
\label{Eq:SUb}
\end{eqnarray}
\begin{equation}
T_{UUbb+UbUb} = 0
\label{Eq:TUb}
\end{equation}

\begin{eqnarray}
S_{DDbb}+S_{DbDb} &=& 16\pi\, \frac{g_{DD}g_{bb}+g_{Db}^2}{{\cal M}^2}\, N\, N_c \left(\frac{2}{9}\, J_{bb}-2\, J_{bb}^\prime\right) J_{DD}\, M_D^2 \nonumber \\
&=& S_0\, \frac{\left(g_{DD}g_{bb}+g_{Db}^2\right)\, N\, N_c}{8\pi^2}
\frac{M_D^2}{{\cal M}^2}\, \left(\frac{2}{3}\, {\overline \log}\frac{{\cal M}^2}{M_b^2}-1\right){\overline \log}\frac{\Lambda^2}{M_D^2} \nonumber \\
&\times&\left[1+{\cal O}\left(\frac{M_Q^2}{\Lambda^2}\right)+{\cal O}\left(\frac{M_t^2}{{\cal M}^2}\right)\right]
\end{eqnarray}
\begin{equation}
T_{DDbb}+T_{DbDb} = 0
\end{equation}

\begin{eqnarray}
S_{qqqq} &=& -16\pi\, \frac{g_{qq}^2}{{\cal M}^2}\, N_c^2 \left[\frac{4\, J_{bb}-2\, J_{tt}}{9}+2\left(J_{tt}^\prime-J_{bb}^\prime\right)\right] J_{tt}\, M_t^2\nonumber \\
&=& -S_0\, \frac{g_{qq}^2\, N_c^2}{12\pi^2}\, \frac{M_t^2}{{\cal M}^2}\, \left[
2\, {\overline \log}\, \frac{{\cal M}^2}{M_b^2} - {\overline \log}\, \frac{{\cal M}^2}{M_t^2}\right]\, {\overline \log}\, \frac{{\cal M}^2}{M_t^2}
\left[1+{\cal O}\left(\frac{M_Q^2}{\Lambda^2}\right)+{\cal O}\left(\frac{M_t^2}{{\cal M}^2}\right)\right]
\end{eqnarray}
\begin{eqnarray}
T_{qqqq} &=& \frac{4\sqrt2\, G_F}{\alpha}\, \frac{g_{qq}^2}{{\cal M}^2}\, N_c^2\left(J_{tt}\, M_t^2\right)^2 \nonumber \\
&=& T_0\, \frac{g_{qq}^2\, N_c^2}{4\pi^2}\, \frac{M_t^2}{{\cal M}^2} \left({\overline \log}\, \frac{{\cal M}^2}{M_t^2}\right)^2
\left[1+{\cal O}\left(\frac{M_Q^2}{\Lambda^2}\right)+{\cal O}\left(\frac{M_t^2}{{\cal M}^2}\right)\right]
\end{eqnarray}

\begin{eqnarray}
S_{tttt} &=& 16\pi\, \frac{g_{tt}^2}{{\cal M}^2}\, N_c^2 \left(\frac{4}{9}\, J_{tt}-2J_{tt}^\prime\right)J_{tt}\, M_t^2 \nonumber \\
&=&S_0\, \frac{g_{tt}^2\, N_c^2}{8\pi^2}\, \frac{M_t^2}{{\cal M}^2}
\left(\frac{4}{3}\, {\overline \log}\frac{{\cal M}^2}{M_t^2}-1\right){\overline \log}\frac{{\cal M}^2}{M_t^2}
\left[1+{\cal O}\left(\frac{M_Q^2}{\Lambda^2}\right)+{\cal O}\left(\frac{M_t^2}{{\cal M}^2}\right)\right]
\end{eqnarray}
\begin{eqnarray}
T_{tttt} &=& \frac{4\sqrt2\, G_F}{\alpha}\, \frac{g_{tt}^2}{{\cal M}^2}\, N_c^2\left(J_{tt}\, M_t^2\right)^2 \nonumber \\
&=& T_0\, \frac{g_{tt}^2\, N_c^2}{4\pi^2}\, \frac{M_t^2}{{\cal M}^2} \left({\overline \log}\, \frac{{\cal M}^2}{M_t^2}\right)^2 
\left[1+{\cal O}\left(\frac{M_Q^2}{\Lambda^2}\right)+{\cal O}\left(\frac{M_t^2}{{\cal M}^2}\right)\right]
\end{eqnarray}

\begin{eqnarray}
S_{bbbb} = 0\, \quad\quad T_{bbbb} = 0
\end{eqnarray}

\begin{eqnarray}
S_{ttbb} &=& -16\pi\, \frac{g_{tt}g_{bb}}{{\cal M}^2}\, N_c^2 \left(\frac{2}{9}\, J_{bb}-2\, J_{bb}^\prime\right) J_{tt}\, M_t^2 \nonumber \\
&=&-S_0\, \frac{g_{tt}g_{bb}\, N_c^2}{8\pi^2}\, \frac{M_t^2}{{\cal M}^2}
\left(\frac{2}{3}\, {\overline \log} \frac{{\cal M}^2}{M_b^2}-1\right) {\overline \log} \frac{{\cal M}^2}{M_t^2}
\left[1+{\cal O}\left(\frac{M_Q^2}{\Lambda^2}\right)+{\cal O}\left(\frac{M_t^2}{{\cal M}^2}\right)\right]
\end{eqnarray}
\begin{equation}
T_{ttbb} = 0
\end{equation}

\begin{eqnarray}
S_{UDtb}=S_{UtDb}=0\ , \quad\quad
T_{UDtb}=T_{UtDb}=0 \ .
\end{eqnarray}

\bibliographystyle{aipnum4-1}

\bibliography{FullETCArxive.bib}

\end{document}